\documentclass[numberedappendix]{aastex}
\usepackage{emulateapj5,epsfig}
\newcommand{\etal}{{et~al.}}
\newcommand{\fx}{ergs s$^{-1}$ cm$^{-2}$}

\newcommand{\lx}{ergs s$^{-1}$}
\newcommand{\lxh}{$h_{50}^{-2}$ ergs s$^{-1}$}
\newcommand{\bggmph}{($h_{50}^{-1}$~Mpc)$^{1.77}$}
\newcommand{\bggmp}{Mpc$^{1.77}$}
\newcommand{\eband}{$0.3-3.5$~keV}
\newcommand{\ebandp}{($0.3-3.5$~keV)}
\newcommand{\hkpc}{$h_{50}^{-1}$~kpc}
\newcommand{\rosat}{{\it ROSAT}}
\newcommand{\figdir}{}

\newcommand{\myfigure}{figurehere}

\newcommand{\mydeluxetable}{deluxetable}
\newcommand{\myacknowledgments}{\acknowledgments}

\makeatletter
\newenvironment{tablehere}
  {\def\@captype{table}}
  {}

\makeatother

\submitted{Received 2000 November 1; accepted 2001 August 29}
\journalinfo{The Astrophysical Journal}

\shorttitle{New X-ray Clusters in the EMSS I}
\shortauthors{Lewis \etal}

\begin{document}

\title{New X-ray Clusters in the {\it Einstein} Extended Medium
Sensitivity Survey I: Modifications to the X-ray Luminosity Function}

\author{Aaron D. Lewis\altaffilmark{1,2,3}, John T.
Stocke\altaffilmark{1,2}, E. Ellingson\altaffilmark{1,2}}

\affil{Center for Astrophysics and Space Astronomy, University of
Colorado, 389 UCB,  Boulder, CO 80309}
\altaffiltext{1}{lewisa@uci.edu, stocke@casa.colorado.edu,
e.elling@casa.colorado.edu}
\altaffiltext{2}{Visiting Astronomer, Kitt Peak National Observatory, National Optical Astronomy
Observatories, which is operated by the Association of Universities for Research in Astronomy, Inc.
(AURA) under cooperative agreement with the National Science Foundation.}
\altaffiltext{3}{Current Address: University of California, Irvine, Department of Physics and Astronomy,
4171 Frederick Reines Hall, Irvine, CA, 92697-4575}

\and
\author{Eric J. Gaidos\altaffilmark{4}}

\affil{Center for Space Research, Massachusetts Institute of
 Technology, Cambridge, MA 02139}

\altaffiltext{4}{Current Address: Division of Geology and Planetary Science, California Institute of
Technology 170-25, Pasadena, CA 91125,
gaidos@gps.caltech.edu}

\received{November 1, 2000}

\accepted{August 29, 2001}

\journalid{564}{1}

\articleid{}{}

\begin{abstract}  
  
The complete ensemble of {\it Einstein} Imaging Proportional Counter (IPC) X-ray images has been
re-processed and re-analyzed using a multi-aperture source detection algorithm. A catalog of 772 new
source candidates detected within the 38 arcmin diameter central regions of the 1435 IPC fields
comprising the Extended Medium Sensitivity Survey (EMSS) has been compiled.  By comparison, 478 EMSS
sources fall within the same area of sky.  A randomly-selected subsample of 133 fields was
examined; 73 sources were detected and compared with 49 original EMSS sources in the same region of
sky. We expect, based on confusion statistics, that most of these sources are either the summation of
two or more lower count rate point sources that fall within the larger detection apertures or are single
point sources. An optical imaging study discovered one possible cluster of galaxies among 43
identified sources, suggesting that $\leq2.3\%$ of the full catalog of sources are extrapolated to be
actual distant ($z\geq0.14$) clusters whose extended X-ray structure kept them from being detected in
the EMSS despite having sufficient total flux. 

We have constructed other subsamples specifically selected to contain those X-ray sources most likely to
be clusters based upon additional X-ray and optical criteria.  Both a database search and an optical
imaging study of these subsamples have found several new distant clusters, setting a firm lower limit on
the number of new clusters in the entire catalog. Given both the numbers of new EMSS clusters and their
spectroscopic or photometric redshifts, we estimate that the original EMSS cluster sample is $72-83\%$
complete.  We update the \citet{hen92_h92} EMSS distant ($z\geq0.14$) cluster sample with more recent
information, and use the redshifts and X-ray luminosities for these new EMSS clusters to compute revised
X-ray Luminosity Functions (XLFs) in the three redshift shells defined by
\citet{hen92_h92}.  The addition of these new high-$z$, high-L$_X$ clusters to the EMSS is sufficient to
remove the requirement for ``negative'' evolution at high-L$_X$ out to $z\sim0.5$. Although the best
estimate of the EMSS XLF at $z=0.3-0.6$ and log L$_X$ of $44.9-45.2$ \lx{} falls $1\sigma$ below the
low-$z$ ($<0.3$) XLF, the optical identification of the full 772 source catalog remains incomplete. We
conclude that the EMSS has systematically missed
clusters of low surface brightness. Since all X-ray cluster surveys are less sensitive to low surface
brightness emission, they may be also be affected.
\end{abstract}


\section{Introduction\label{sec_intro}}

The {\it Einstein} Extended Medium Sensitivity Survey (EMSS) of X-ray sources was constructed from
serendipitous detections in Imaging Proportional Counter (IPC) images at high Galactic latitude free of
very bright or extended target sources \citep{gio90a}. This catalog has served as the foundation for
investigations of the statistical properties of several different classes of objects, including stars,
active galactic nuclei (AGN), BL Lac objects, and clusters of galaxies
\citep{sto91,gio94}.  The sample of EMSS X-ray clusters is of special cosmological interest since these
objects can be detected to high redshift and, because of the density-squared dependence of the X-ray
emission and the low volume filling factor ($\sim 10^{-7}$) of X-ray luminous clusters,  X-ray-selected
samples are not significantly affected by the confusion and projection effects which plague
optically-selected catalogs. Thus X-ray cluster surveys are potentially powerful tests of cosmology and
structure formation models \citep[e.g.,][]{eke96,don98,bah98,don99a}.  Our knowledge of X-ray cluster
statistics and evolution has dramatically increased with the advent of large solid-angle, high-sensitivity
X-ray surveys such as the EMSS 
\citep[H92 hereafter]{gio90a,hen92_h92}, the {\it ROSAT} All Sky Survey
\citep[RASS,][]{tru93,ebe96,ebe01,deg99,mos00}, as well as the investigation of the deepest regions of
the RASS, the North Ecliptic Pole Survey \citep[NEP,][]{gio01}, and collections of serendipitous sources
found in deeper {\it ROSAT} pointed observations
\citep{ros95,sch97a,vik98a,rom00,per00}.

The interpretation of these surveys as tests of cosmological models must include various theoretical
effects, the most pronounced of which may be the thermal history of the X-ray-emitting gas
\citep{evr91,kai91}. In addition, there may be systematic errors introduced by the methods used to analyze
the X-ray imaging data.  The possibility that past X-ray source catalogs developed from {\it
 Einstein} IPC images suffer from bias or incompleteness motivated re-analyses of the data
\citep[OHG97 hereafter]{ham93,mor96,opp97_ohg97}. Although the {\it Einstein} data have been surpassed by
the increased sensitivity and larger field of view of the {\it ROSAT} Position Sensitive Proportional
Counter (PSPC),  the EMSS catalog remains one of the largest sky-area sample of X-ray clusters against
which others are compared. Due its large areal coverage, it is the only published survey to explore the
highest luminosity region of the cluster X-ray Luminosity Function (XLF), in which the greatest leverage is
obtained for constraining evolution in the XLF.  Further, the EMSS catalog is one of the few
large-sky-area source catalogs which has been thoroughly investigated and almost completely identified
optically regardless of inferred source character; e.g., some of the deep {\it ROSAT} cluster searches have
used algorithms to select only extended X-ray emitters, which are therefore the only sources investigated
optically \citep[c.f.,][]{gio01}.   Additionally, accurate X-ray temperatures of high$-z$ EMSS clusters
obtained by {\it ASCA}
\citep{hen97,don98,don99b} have allowed significant constraints to be placed on the value of
$\Omega_{\rm matter}$ \citep{bah98,don99a,voi00,hen00}. With such a wealth of unique data now available
for the EMSS sample, re-visiting the IPC data is still warranted.

New work on the IPC imaging data is characterized by two advancements. First, the IPC response
flat-fielding, temporal event filtering, and exposure maps have been improved, reducing systematic errors
that limit source detection (for detailed discussion, see OHG97).  Second, a more sophisticated algorithm
for source detection has been implemented.  Previous IPC source catalogs such as the EMSS were constructed
using a single 2.4 arcmin square detection aperture whose size was optimized for the detection of point
sources.  Thus such a catalog is biased towards point sources or sources with a high central surface
brightness.  While such a catalog may be complete to a certain X-ray flux {\it within the detection cell},
more extended sources with integrated fluxes above the survey threshold levels may have been excluded due
to their lack of concentrated emission.  Since X-ray clusters are often resolved in imaging data, this
bias could significantly affect their apparent statistics in the catalog.
\citet{pes90} noted that up to 75\% of the clusters in the original {\it Einstein} Medium Sensitivity
Survey would have been excluded had they lacked centrally-peaked X-ray emission. But it is now known from
{\it ROSAT} imaging that not all EMSS clusters are so centrally-peaked as to suggest ``cooling flows''
\citep[e.g.,][]{don92,L99} and some are quite diffuse as X-ray emitters
\citep[e.g., MS 1621.4+2146;][]{mor98}.  Using simple models of the X-ray surface-brightness distribution
of clusters, H92 and
\citet{don92} estimated that the fraction of emission of a typical EMSS cluster falling outside the 2.4
arcmin square detection cell was between 38 and 93\%, with a median value of 58\%. For this reason, the
XLF analysis by \citet{gio90b} and H92 used only EMSS clusters at $z\geq 0.14$, where the percentage of
the total flux falling outside the detect cell was not prohibitively large. Furthermore, nearby ($z\leq
0.14$) clusters were frequently the targets of the IPC observations, and were therefore unavailable to the
EMSS as serendipitous sources.

This work addresses the question of whether there are a significant number of previously undetected
serendipitous sources which fall in the EMSS fields and have count rates sufficiently high to justify
inclusion in the EMSS catalog, but which have been previously excluded due to systematic errors or biases
in the construction of that catalog. OHG97 constructed an X-ray source catalog from a complete database of
archived IPC images using a detection algorithm that employed multiple apertures of different sizes to
minimize X-ray surface-brightness selection effects.  Several thousand new unidentified sources were
found, of which more than 300 appear to be significantly extended. The OHG97 IPC source catalog is used as
the starting point for this investigation, but several important observational details in the selection
process must be considered herein in order to match the EMSS selection as accurately as possible,
excepting detection aperture. While OHG97  eventually selected sources based on angular extent, all
potential sources within the EMSS fields are considered here.

In \S \ref{sec_const} the OHG97 catalog construction method is summarized and the sources scrutinized to
determine if each would have been included in the EMSS, in an attempt to make a truly flux-limited
catalog. In \S \ref{sec_invest} the nature of these ``new'' X-ray sources is investigated using available
databases and optical observations of a randomly-selected subsample of these new sources. Since very few
new distant clusters were found in the random sample, in \S \ref{sec_nonrandom} we investigate two
non-random subsamples of sources specifically selected to maximize the discovery of distant clusters. 
Using the results from these subsamples, the source catalog is finalized in \S \ref{sec_impl} for the most
accurate comparison with the EMSS as possible; i.e., a sample  whose selection matches the EMSS selection
process as accurately as possible, given the different methods. The statistical effect of these new
sources on the EMSS XLF results is also estimated in \S \ref{sec_impl}. A summary and conclusions are
presented in \S
\ref{sec_concl}. We assume $H_0 = 50 h_{50}^{-1}$ km s$^{-1}$ Mpc$^{-1}$, $\Omega_0 = 1.0$, and $\Lambda =
0$, unless otherwise stated.
 
\section{Construction of the New Source Catalog\label{sec_const}}

The Imaging Proportional Counter (IPC) was one of the primary instruments on the High Energy Astrophysical
Observatory (HEAO)-2 {\it Einstein} satellite which operated for two and a half years beginning in
November 1978 \citep{gia79}. The IPC was an imaging proportional counter and the X-ray telescope/detector
combination was sensitive to photons in the energy range 0.2 to 4.0 keV with an effective area of about
100 cm$^{2}$. The field of view was 76 arcmin on a side with an on-axis resolution of 1.5 arcmin. During
its operation, the IPC obtained nearly 4100 images with exposure times ranging from 100-56,000 seconds.
The celestial coordinates of every photon detected in IPC images are recorded on optical disk archives at
the Columbia University Center for Astrophysics. This merged image database was used to construct a new
IPC source catalog independent of the EMSS. The {\it Einstein} data products were also processed by the
CfA at Harvard, and are available through the HEASARC service\footnote{A service of the Laboratory for
High Energy Astrophysics (LHEA) at NASA/ GSFC and the High Energy Astrophysics Division of the Smithsonian
Astrophysical Observatory (SAO)}.

\subsection{Source Detection\label{subsec_detect}}

The source detection process has been described in detail in
\citet{ham91}, \citet{mor96} and OHG97. Here, only the essential elements of the algorithm are reviewed.
Briefly, the distinctions between our procedures and the standard {\it Einstein} processing system include
the application of a flat field to compensate for IPC detector nonuniformities, improved data editing, a
circular (rather than square) source aperture, a local background determination for each search cell, and
an iterative source search algorithm using apertures of different sizes.

The preliminary source catalog was constructed using only the unobstructed $38 \times 38$ arcmin center of
the IPC field of view. Cumulative IPC event lists, exposure maps, and energy-dependent flat fields were
constructed for each spacecraft orbit, with restrictions on allowable energy channels and telescope-Sun
angles. The computed count rates in the $0.3-3.5$~keV bandpass for all orbits were then summed into
cumulative count and count-rate maps with 32 or 64 arc-second pixels. The maps were scanned with a series
of four circular apertures with different diameters: 2.5, 4.7, 8.4, and 12.2~arcmin (alternatively, we
will refer to these as the 1st, 2nd, 3rd, and 4th apertures, respectively). The smallest aperture is the
optimal size for the detection of point sources and has 85\% of the sky area of the 2.4 arcmin square
detect cell aperture of the EMSS. The size of the largest aperture is limited  by the field of view.
Subsequent analysis showed that, for the 12.2 arcmin aperture, flat-fielding and vignetting corrections
are too large and source confusion too frequent to make detections in that aperture viable. Therefore, the
12.2 arcmin aperture data was not used as a sole criterion for source detection. The background was
estimated in a circular annulus  surrounding each detection aperture with an area of 85 arcmin$^2$. For
the smallest aperture, the annulus was the area between 3 and 6 arcmin in radius, and for each larger
detection aperture the other background annuli were scaled up appropriately . The Poisson noise from both
the source aperture and background annulus was also calculated.  An initial scan of the data was made to
find all possibly significant detections with a ratio of signal to Poisson noise exceeding 2.5 and a
sufficient fraction of reliable pixels in the source and background apertures (roughly 60\% and 30\%,
respectively). These fractions are less restrictive than the EMSS algorithms of necessity, since the
larger detection apertures require that a larger fraction of potential sources are close to rib support
structures and other, brighter sources. Detection proceeded iteratively over the entire sky map, with the
threshold count-rate decremented from an initial high value down to the 2.5$\sigma$ minimum. Pixels
associated with detected sources at each iteration were masked out for successive iterations.  The entire
sky map was analyzed separately with each aperture.

A catalog of sources was then constructed from the three useful lists of detections, by matching them if
the center of one detection fell within the aperture of another. A single source usually consists of
multiple detections in each aperture. However, only one detection (with the highest S/N) was retained for
each aperture and the others were discarded. A source was retained in the catalog only if there was a
$4\sigma$ detection in at least one aperture. The location of each source was defined to be the centroid
computed as the weighted average of all the aperture centers
$\vec{x_i}$;
\begin{equation} \vec{X_c} = \frac{\Sigma_i
\vec{x_i}(R_i/\sigma_i)^{-2}}{\Sigma_i (R_i/\sigma_i)^{-2}},
\end{equation} where $R_i$ is the aperture size and $\sigma_i$ the signal-to-noise in that aperture.  This
is only an approximation since it assumes that the measurements in the different apertures are independent,
which they are not.   The single-standard deviation uncertainty in the position is taken to be
$\sim min\left[R_i/\sigma_i\right]$, and we note that all source positions are accurate to better than one
arcmin, with an average uncertainty of 28 arcsec.

A total of 7419 sources were identified in this manner. To eliminate contamination by Galactic emission or
known sources of diffuse emission, only sources at Galactic latitude $|b| > 20^{\circ}$ were retained.
Also, any sources within 5$^{\circ}$ of the Large Magellanic Cloud, 2.67$^{\circ}$ of the Small Magellanic
Cloud, 2.67$^{\circ}$ of the Coma Cluster, or 1.6$^{\circ}$ of Messier 31 were excluded. At this stage, 
the IPC source list contained 6600 sources.

\subsection{New Sources in EMSS Fields\label{subsec_new}}

The original EMSS catalog of 835 X-ray sources was compiled from serendipitous sources detected in 780
deg$^2$ within 1435 separate IPC images at $|b| > 20^{\circ}$ \citep{gio90a}. To construct a list of new
sources which can be directly compared with the EMSS, the 6600 IPC sources were culled to include only
those sources which have measured count-rates above the required threshold for the corresponding field
(i.e., sufficient to generate $\sigma > 4$ in the EMSS detect cell if the counts were concentrated in a
point source profile), and those sources which fall within EMSS fields. These restrictions result in a
catalog of 979 sources. It was also necessary to cull some of these 979 sources from the catalog because
they would {\bf not} have been included in the EMSS for other specific reasons described below.

To estimate the S/N that each source would have had in the EMSS detect cell, the
source and background counts in the EMSS field must be predicted by including several factors:

To estimate count rates in the EMSS detect cell, the effective exposure time at the location of a source
in each IPC image was calculated by adjusting the ``live'' time at the image center for large angle
scattering by the X-ray telescope mirror (a factor of 0.847) and vignetting. The vignetting function
$V(\theta)$ derived by
\citet{har90} for 1.5 keV energies was adopted. As a function of the off-axis angle
$\theta$ in arcmin it is given by
\begin{equation} 
V(\theta) = 0.997 - 8.25\times10^{-3}\theta - 3.125\times10^{-4}\theta^2, 
\end{equation}
for $\theta < 12$ arcmin, and
\begin{equation} 
V(\theta) = 1.1049 - 0.02136\theta, 
\end{equation}
at larger angles. At the 19 arcmin limit of each subfield, the vignetting correction is 0.70.

Second, the IPC background was assumed to be uniform in the vicinity of each source and the number of
background counts was corrected by the ratio of the area of the EMSS detection aperture (5.76 arcmin$^2$)
to that of the aperture used in the actual detection. Additionally, the predicted total number of EMSS
source counts calculated above was multiplied by the fraction of the point response function inside the
EMSS detection cell (a factor of 0.885).  A similar correction was {\it not} applied to the count-rates in
the new apertures but that correction is quite small. Using the estimated EMSS detect cell source and
background counts, we determined the significance of each detection. The S/N of each source was calculated
using the formula 
\begin{equation}
\sigma = \frac{C_S}{\sqrt{C_S + C_B}}, 
\end{equation}
where $C_S$ is the source counts and $C_B$ the background counts. The calculation was performed using the
source and background counts for each of the three smallest apertures corrected as described above. This
calculation determines whether the newly discovered source would still be greater than $4\sigma$ in the
EMSS catalog if a different aperture size had been used in the EMSS detection algorithm. We excluded any
sources from the catalog that did not meet this requirement.

We then scrutinized the catalog to eliminate any remaining sources which should not be included in the
EMSS. Only sources falling within specific annuli centered on each of the EMSS field centers were
considered. The outer annulus border is a circle 19 arcmin in radius which lies just inside the IPC
detector ribs and includes the most sensitive detection area.  Within this radius the vignetting
correction is less than 30\% and is nearly unaffected by the satellite roll angle.  The inner border is 6
arcmin in radius and preferentially screens sources which were the targets of pointed observations rather
than serendipitous detections.  Although this region is different than that used to construct the EMSS (a
square 45 arcmin on a side with the central 5 arcmin removed), the restrictions can be applied {\it post
facto} to the EMSS catalog to generate a subset of the EMSS catalog which can be directly compared to the
new source list.  To remove any overlap with the EMSS catalog, all IPC sources within 6 arcmin of any EMSS
source are removed and, to remove duplications, all catalogued sources within 6 arcmin of another source
are also excluded. Fifty-four such targets were identified and removed.  

While the 19 arcmin maximum radius from the IPC center was imposed to avoid shadowing by the detector
``ribs'', the IPC center is actually offset with respect to these ribs and, moreover, some sequence number
frames are actually combinations of sub-exposures taken at the same sky location but at different
telescope roll angles. The result of these multiple exposures for the current analysis is that the rib
pattern rotates with respect to the field. The combination of these two effects means that some sources at
a radius $r < 19$~arcmin away from the field center still have non-zero rib-edge codes (i.e., vignetting
by the IPC rib structure or the edges of the detector area are quantified by a ``rib-edge'' code,
\citealt{har90}; we have used the same rib-edge code criteria to eliminate sources as that described in
\citealt{gio90a}) and so would not have been included in the EMSS source catalog. To remedy this situation
we have scrutinized individually for rib-edge codes (a) all sources with
$16<r<19$ arcmin and (b) all sources detected in sequence numbers made up of multiple sub-exposures taken
at different roll angles.  Thirty-eight sources were excluded because they would also have been excluded
from the EMSS due to shadowing by the ribs.

In some cases, due to a poorly known target position, the desire to observe other nearby sources in the
field-of-view of the IPC or simply some other circumstance (e.g., mispointing), the target of the IPC
observation was not within the 6 arcmin central region. Also some targets are quite large in angular size
(e.g., M101, NGC 253, etc.) and so occupy more than 6 arcmin within the IPC field. We have eliminated both
the specific mispointed targets of IPC observations, as well as sources which are related to the target
(i.e., within the optical extent of very extended targets).  Forty-eight sources were identified as the
mispointed target of the observation and eliminated; 12 additional sources were eliminated for being
related to very extended targets.

In addition, some regions of the sky were observed more than once by the IPC detector. In some cases, the
additional exposures were fields {\it not} included in the original EMSS. For these fields, the S/N of any
detected, non-variable source will be increased over the value that would be measured by using only the
EMSS exposure time. We have individually investigated every set of merged fields (overlapping with the
central coordinates of the fields coincident to $<1$~arcsec) in the catalog. We identified those sources
whose S/N would be reduced to less than 4 in the first three apertures if only the EMSS exposure time was
used. We eliminate 55 such sources.

Finally, there exist multiple IPC exposures in some fields which overlap but are not coincident, thereby
increasing the exposure time for only the overlapping regions. If the overlapping observation was not
included in the original EMSS, a detected source will have an increase in measured S/N similar to the
merged field scenario described above. The IPC detector area is so large that the number of such
occurrences in the catalog is prohibitively large to investigate individually.  It is also difficult to
estimate the true effective exposure time in such overlapping regions because of the large variation in
vignetting 
across the IPC field. In \S \ref{subsec_moreacc} we make a
statistical estimate of the effect of these occurrences on the catalog due to overlapping non-EMSS
observations.

Excluding sources for all of the reasons cited above (except the statistical estimate for multiple
exposures to be applied later) results in a catalog containing 772 sources.  An electronic version of the
772 source catalog can be obtained from the first
author\footnote{\url{http://casa.colorado.edu/$\sim$lewisad/research/nemsscatalog.html}}. 

While the new IPC sources are 772 in number, there are only 478 (of 835 total) original EMSS sources
falling within the same restricted survey area. Figure~\ref{fig_cr} compares the $0.3-3.5$~keV count rates
calculated in the 2.5 arcmin aperture to the original EMSS values for 334 of these 478 sources. We have
compared only sources thought to be point-like on the basis of their optical identifications (i.e., EMSS
sources identified as either stars, BLLacs, or AGN; any EMSS source which is even possibly a combination
of two or more sources has also been eliminated from this plot, see also a similar plot in OHG97).  The
agreement is generally very good and, significantly, there are not a substantial number of sources with
OHG97 count rates significantly {\it higher} than EMSS values.  Thus new sources are not being included
due to some systematic upwards bias in the re-estimated count rate.
\begin{\myfigure}
\centerline{\epsfig{file=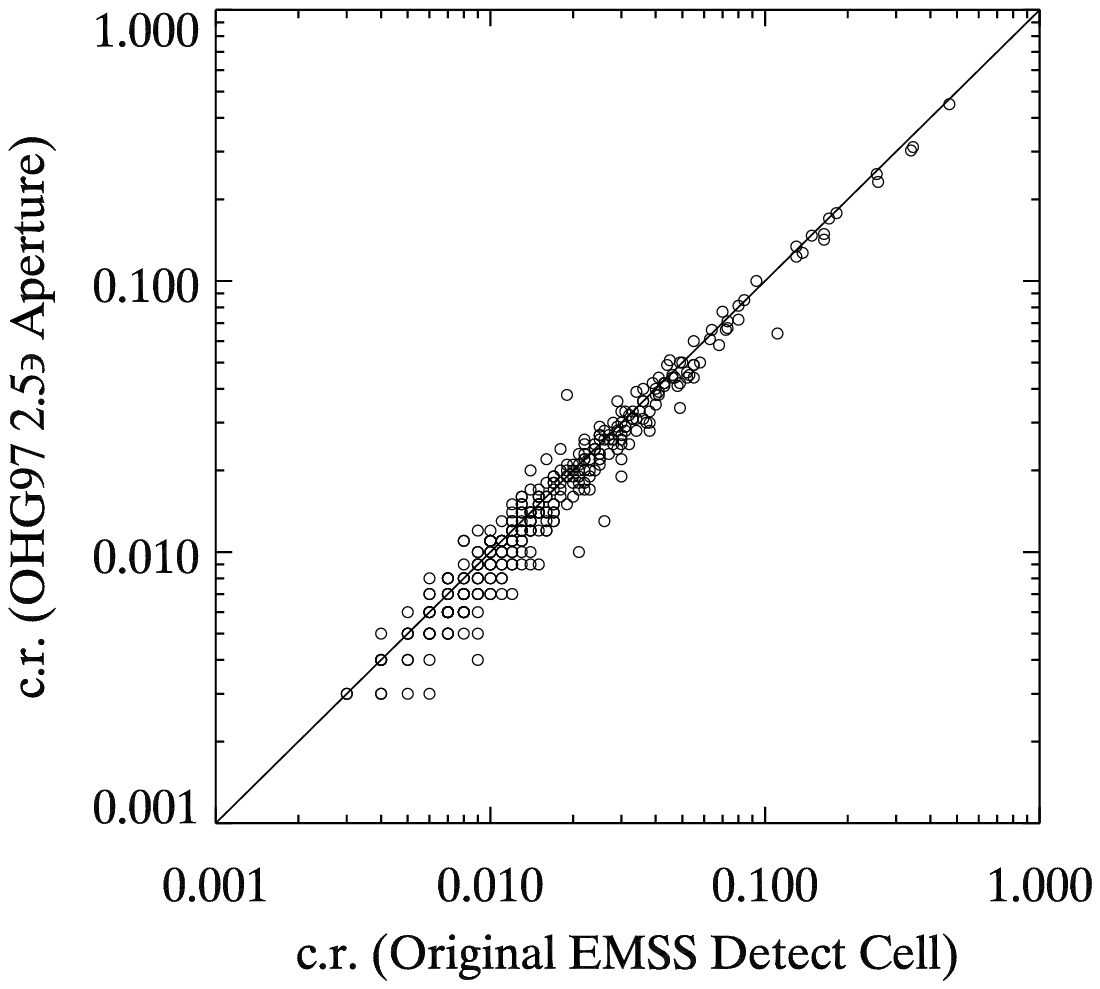,scale=0.6}}
\caption[Comparison of Count Rates in Aperture 1 vs. the EMSS Detect Cell]{Comparison of the $0.3-3.5$~keV
count rates (c.r., in units of counts s$^{-1}$) estimated in the 2.5 arcmin aperture of the OHG97 catalog
to those estimated by the EMSS
\citep{gio90a} for 334 point-like EMSS sources within the central region of the EMSS fields. The solid
line represents equivalence between the two estimates.
\label{fig_cr}}
\end{\myfigure}
\smallskip
\epsscale{1.0}

\subsection{Estimates of Source Confusion\label{subsec_est}}

The use of apertures larger than optimal for point-source detection suggests the possibility that spurious
detections may occur if more than one unrelated source with flux below the detection limit appears in the
same aperture.  The ``excess'' variance (over other noise sources) produced by point-to-point fluctuations
in numbers of undetected sources in a beam has been traditionally exploited to constrain number counts
below the detection threshold \citep[e.g.,][]{ham87}. Here, the hypothesis that such variations are
responsible for a significant number of these new IPC sources is evaluated; i.e., the probability that
more than one source is present in the detection  aperture, which when combined obtains a count rate above
the detection threshold.

A method that can be used to estimate the number of multiple sources in any one detection cell as a
function of detected source flux is described in \citet{sto91}. Given the detection aperture area and a
typical source flux for EMSS sources, S.L. \& S.D. Morris
\citep[in][]{sto91} developed a probabilistic formalism which predicted that
$\sim16$  of the EMSS sources with $f_{0.3-3.5}\geq2 \times 10^{-13}$ \fx{} were confused and were
actually two sources of lower flux, which combine to make up the detected count rate. By ``confused'' it
was meant that the second, fainter source contributed
$\geq20\%$ of the total flux in the detect cell (and thus contributed significantly to a 4$\sigma$ source
near the detection limit). The predicted number of confused sources ($16\pm4$) was verified
observationally in the EMSS by the number of X-ray source error circles which contained more than one
plausible optical identification (ID); i.e., more than one plausible X-ray emitter based upon their
optical properties was found for 10 sources at this flux limit and above. A few other cases have been
found since that time based upon reobservation of EMSS sources using the higher spatial resolution of the
{\it ROSAT} PSPC or HRI.  \citeauthor{sto91} noted that, based upon their high surface densities, QSOs and
stars are the most likely confusing source populations (i.e., pairs or multiples of point sources).

Using the scaling relations for source confusion with flux and aperture size given above,  the EMSS
baseline value for confusion can be extrapolated to the OHG97 methodology (the accuracy of which has been
verified by the optical ID work of the EMSS team).   Table \ref{tab_confusion} shows the results of these
calculations, the percentages of preliminary catalog sources which are ``confused''.  Percentages are
shown for the highest count rates encountered in the catalog, for the median count rate, for a low value
of the count rate and for the lowest count rates encountered in the catalog  as a function of the various
detect cell aperture sizes used herein.  Several conclusions can be drawn immediately: (1) as stated for
other reasons before, the 12\farcm2 aperture is not usable for making independent detections, being so
large that virtually all but the brightest  detections in that aperture are confused; (2) even the
8\farcm4 aperture has a large fraction of confused sources at the fainter fluxes in the catalog and so
sources detected only in that aperture must be viewed skeptically;  (3) virtually all of the sources near
the faint flux limit of the catalog are confused in all but the smallest aperture. Indeed, the values in
Table
\ref{tab_confusion} suggest that few sources below count rates of 0.01 are not confused. For reference, we
show in Figure \ref{fig_crhistogram} the distribution of count rates for the entire 772 source catalog. We
note that sources whose S/N was less than 2.5 in a particular aperture will not have a corresponding count
rate value included in this histogram. Referencing Table \ref{tab_confusion}, we can see from Figure
\ref{fig_crhistogram} that there is a significant population of higher count rate sources which are not
expected to be confused, particularly in apertures 2 and 3 at $> 0.02$, $0.03$ counts s$^{-1}$,
respectively.

\begin{tablehere}
\begin{center}
\caption{Estimated Percentages of ``Confused'' Sources\label{tab_confusion}}
\begin{tabular}{crrrr}
\tableline
\tableline
Aperture Size   & \multicolumn{4}{c}{Count rate (cts s$^{-1}$)} \\
(arcmin)        & 0.05 & 0.025 & 0.01 & 0.004 \\
\tableline
2.5             & 0.5\%  & 1.8\%   & 12\%   &  50\% \\
4.7             & 1.6\%  & 6.6\%   & 41\%   & 100\% \\
8.4             & 8.3\%  & 21\%    & 100\%  & 100\% \\
12.2            & 11\%   & 45\%    & 100\%  & 100\% \\
\tableline
\end{tabular}
\end{center}
\end{tablehere}
\smallskip 

\begin{\myfigure}
\centerline{\epsfig{file=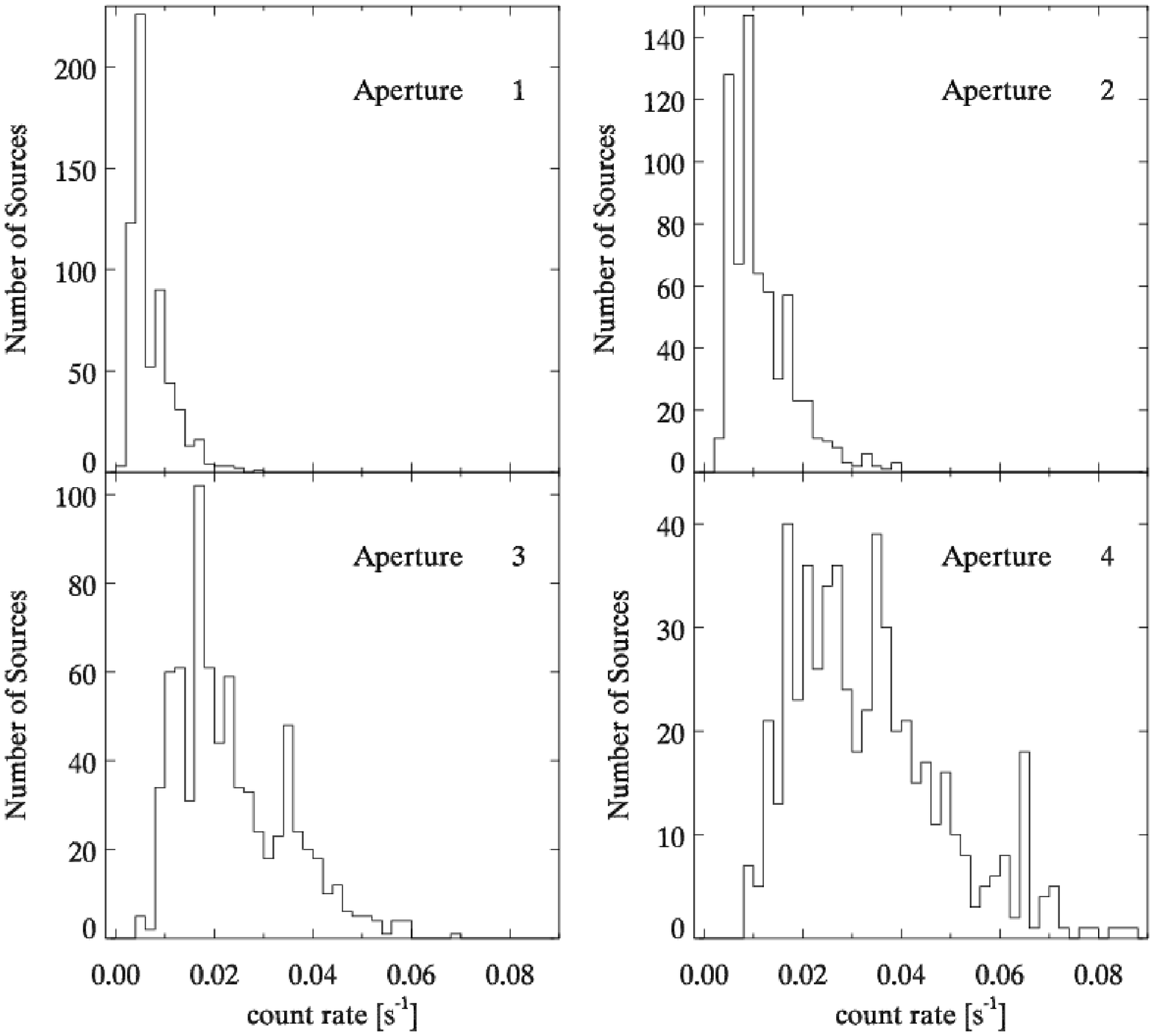, scale=0.40}}
\caption[Distribution of the IPC Count Rates in the 4 Apertures]{Distribution of the measured IPC count
rates (in counts s$^{-1}$) for the 772 source catalog in each of the four apertures used in this work.
Assuming a source spectrum of a 6~keV Raymond-Smith plasma with
$n_H=3 \times 10^{20}$ cm$^{-2}$, a source with count rate of 0.01 counts s$^{-1}$ corresponds to an
unabsorbed flux of $3.3 \times 10^{-13}$~\fx{} in the \eband{} band.
\label{fig_crhistogram}}
\end{\myfigure}
\smallskip
\epsscale{1.0}
 
\section{Investigation of a Random Subsample of New IPC Sources\label{sec_invest}}

The putative new sources in a randomly-selected subsample of 133 of the 1435 EMSS IPC fields were examined
in detail and compared with the original EMSS sources in those same fields. This provides an initial
sample of practical size with which to scrutinize the reality of these sources and to determine their
X-ray nature and identity. This sample size also facilitates a subsequent optical imaging study to
determine whether clusters of galaxies are present among these sources.

\subsection{Random Subsample Properties \label{subsec_random}}

The subsample fields were randomly selected without regard to sequence number, location on the sky, or
exposure time, and thus can be considered representative of the entire collection of EMSS fields. The IPC
sequence numbers of the observations are listed in Table \ref{tab_fields}. These fields include 73 new
serendipitous sources and 49 original EMSS sources. Only 59 of the 133 fields contain even a single new
source. Table
\ref{tab_randomsample} lists the basic X-ray properties for 74 sources\footnote{Source \# 4359, though
included in Table \ref{tab_randomsample}, has been eliminated from the random sample, resulting in 73
random sample sources. See the footnote to Table \ref{tab_randomsample}, and discussion of the source in \S
\ref{subsec_databasesearch}.} including (by column): (1) the catalog number of the source; (2-3) RA and
DEC in J2000 coordinates; (4) the IPC sequence number in which the source was detected;  (5-7) the source
count rates in the 2.5, 4.7, \& 8.4 arcmin diameter apertures respectively (in units of 10$^{-3}$ cts
s$^{-1}$ in the \eband{} band); (8-10) the source statistical significances in these same three apertures
in $\sigma$s; and (11) our final evaluation of the field, (C = definite cluster; C: = possible cluster; X
= definite non-cluster; unmarked objects are unidentified due to a lack of information) based on
information from the literature, X-ray databases, and our observing campaign, all described below.
\begin{table*}
\begin{center}
\caption{Sequence Numbers of a Randomly-Selected Subset of EMSS IPC Fields \label{tab_fields}}
\begin{tabular}{rrrrrrrrrr}
\tableline
\tableline
  207 &   305 &   421 &   443 &   444 &   454 &   470 &   478 &   481 &   498 \\
  500 &   505 &   797 &   852 &   863 &   889 &  1810 &  1937 &  2014 &  2030 \\
 2074 &  2082 &  2101 &  2127 &  2222 &  2602 &  2638 &  2716 &  2720 &  2911 \\
 3018 &  3070 &  3105 &  3176 &  3256 &  3263 &  3368 &  3453 &  3454 &  3471 \\
 3472 &  3530 &  3550 &  3816 &  3984 &  3988 &  3989 &  3993 &  4002 &  4037 \\
 4059 &  4147 &  4250 &  4261 &  4453 &  4499 &  4546 &  4577 &  4606 &  4621 \\
 4946 &  4972 &  5115 &  5125 &  5129 &  5191 &  5259 &  5388 &  5393 &  5397 \\
 5475 &  5504 &  5547 &  5652 &  5666 &  5670 &  5705 &  5708 &  5717 &  5796 \\
 5801 &  5929 &  6311 &  6317 &  6344 &  6407 &  6449 &  6646 &  6694 &  6728 \\
 6733 &  6746 &  6809 &  6835 &  6879 &  7036 &  7116 &  7165 &  7181 &  7204 \\
 7208 &  7426 &  7569 &  7582 &  7605 &  7636 &  7642 &  7668 &  7712 &  7719 \\
 7770 &  7771 &  7801 &  7803 &  7917 &  7957 &  7987 &  8047 &  8332 &  8334 \\
 8385 &  8433 &  8438 &  8439 &  8455 &  8458 &  8464 &  8740 &  8780 &  8838 \\
 8957 & 10382 & 10549 &       &       &       &       &       & &  \\
\tableline
\end{tabular}
\end{center}
\end{table*}

Based on the S/N values in the smallest aperture, very few of the new sources in the random subsample
would be expected to be included in the construction of the original EMSS catalog. To check this we have
measured the S/N of the sources in Table \ref{tab_randomsample} in the 2.4 arcmin square detection
aperture of the EMSS, based on the count rates in the 2.5 arcmin circular aperture used here (see \S
\ref{subsec_new}).  Because of the similar size of the two apertures these estimates should be very close
to the actual values. (If the source was not detected within the 2.5 arcmin aperture, the signal within
the smallest aperture with a detection was used). As expected, the vast majority (97\% of detections in
the smallest aperture) of sources have a S/N less than four in the EMSS detect cell and would not have
been included in the EMSS. They are included here only because larger-sized apertures are used. A few
sources would have had a S/N in the EMSS detect cell only marginally above four, yet they are not found in
the original EMSS, suggesting that the exact size and shape of the detection {\bf and} background
apertures (when combined with small number statistics) can make small differences in the statistical
significance of some sources. 

Do the new sources found in the EMSS fields tend to appear more extended than the original EMSS sources?  A
larger apparent angular extent may be due to a single, extended source, a combination of two or more
confused sources, or to a significantly softer X-ray spectrum, and would partially explain the failure to
detect and include these sources in the original EMSS catalog. The X-ray surface-brightness distribution
of a source can be qualitatively characterized by calculating the ratios of the count rates between the
four different detection apertures used here. The apertures can be combined in three independent ratios:
Here the fluxes in the three larger apertures were each divided by the flux in the next smaller aperture. 
These dimensionless ratios typically have values larger than unity as the larger aperture will tend to
capture more flux than the smaller, but can occasionally be less than one due to Poisson noise or if there
is a significant offset between the locations of the apertures. In Figure \ref{fig_fluxratio} the
normalized distributions of flux-ratio for the 772 putative new catalog sources are compared to those of
the 478 original EMSS sources within the the same inner regions of all 1435 IPC fields. These plots
clearly demonstrate that the new sources tend to appear more extended than the original EMSS sources as
measured by the ratios of the three smallest apertures. The Kolmogorov-Smirnov test of the null hypothesis
that the two data sets are drawn from the same population yields probabilities of
$1\times 10^{-39}$, $1\times 10^{-46}$, and 0.31 using the three respective ratios,
f$_{4\farcm7}/$f$_{2\farcm5}$, f$_{8\farcm4}/$f$_{4\farcm7}$, and f$_{12\farcm2}/$f$_{8\farcm4}$. The
large K-S probability for the third ratio is an additional reason for excluding detections made 
only in the largest aperture. Otherwise, the very small
K-S probabilities for the first two ratios support the hypothesis that some of the new sources were not
originally detected at a S/N $>4$ due to an apparent lower X-ray surface-brightness, but it should be
pointed out that there are substantial overlaps between the flux ratio distributions and this cannot
explain  all of the new sources.

We conclude that these new sources should not have been and were not included by the EMSS because they
have too low a S/N in the EMSS detect cell, but are nevertheless detected with S/N $>4$ using a different
algorithm (i.e., the EMSS did not exclude these sources through some analysis mistake). The most likely
causes for the rejection of these sources using the standard EMSS procedure is that they are either single
extended sources, or collections of two or more point sources of lower individual flux. But, it is also
possible that some point sources just below the 4$\sigma$ detection limit in the EMSS rise to slightly
above the detection limit in one of the new detect cells due to the difference in size and shape of the
detection and background apertures, or the improvement in background determination.

\begin{\myfigure}
\centerline{\epsfig{file=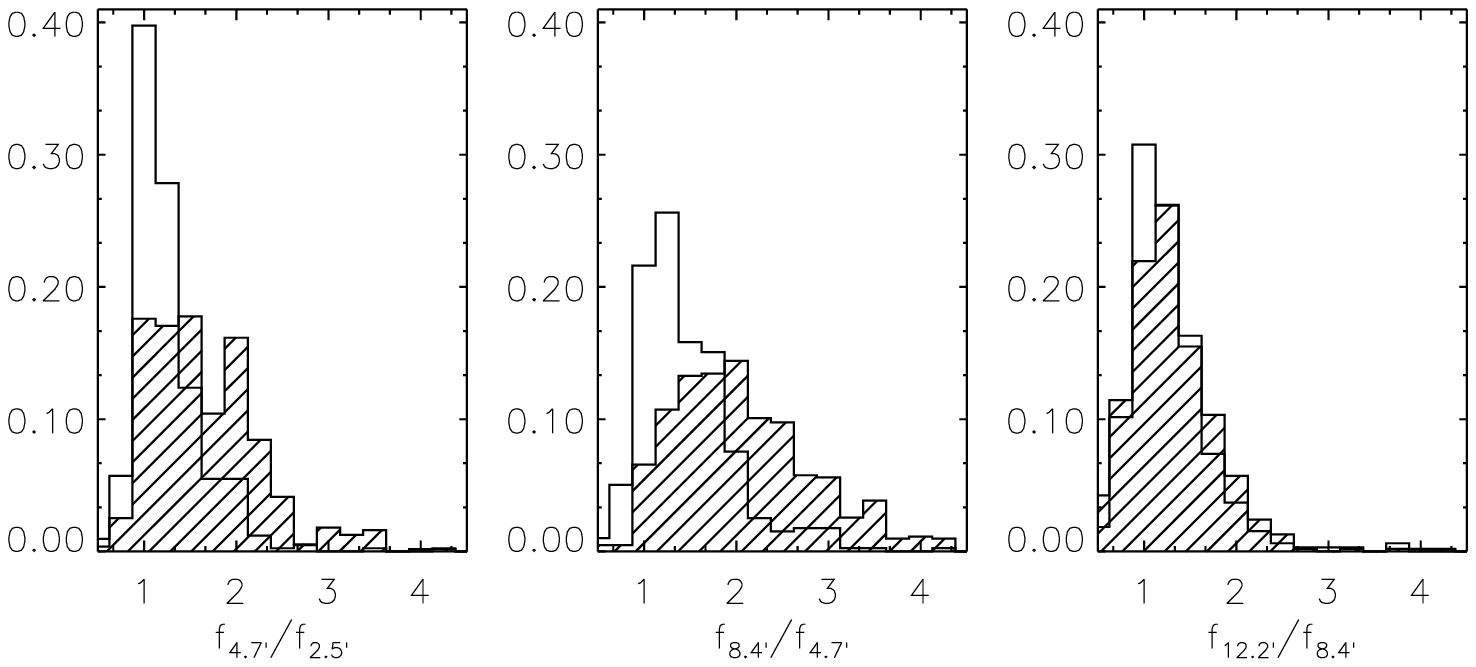, scale=0.55}}
\caption[Distribution of the Flux Ratios for EMSS Sources vs. New Sources]{Normalized distribution of the
ratio of fluxes in different IPC source-detection apertures for original EMSS sources (open histogram) and
new (shaded histogram) source candidates in EMSS fields. Each histogram presents the ratio of successive
pairs of apertures used to generate the catalog.  Larger ratio values indicate that a source is more
extended. In the first two histograms  there is a clear statistical tendency for the new sources to have
higher flux ratios than the original sources.
\label{fig_fluxratio}}
\end{\myfigure}
\smallskip
\smallskip
\epsscale{1.0}

\subsection{Database Search\label{subsec_databasesearch}}
 
Exhaustive searches of the NASA-IPAC Extragalactic Database
(NED\footnote{\url{http://nedwww.ipac.caltech.edu/}}) and the Stellar Information Database
(SIMBAD\footnote{\url{http://simbad.u-strasbg.fr/Simbad}}) were performed to search for plausible
counterparts to each of the 73 new X-ray sources. Objects that are known sources of X-ray emission (i.e.,
clusters or groups of galaxies, AGN, nearby galaxies, and bright or double stars) within 5 arcmin of the
X-ray position (and so potentially contributing to the X-ray flux in one or more of the three smallest
apertures) were noted. Only 11 sources have plausible identifications; including 1 galaxy group at
$z=0.018$, 3 QSOs or AGN, 5 stars, and 2 IRAS galaxies, which are possible AGN.   These sources are listed
in Table \ref{tab_nonxraydbrandom}, along with the references to these IDs, each of which has been checked
for plausibility using the ($f_x/f_V$) method of \citet{sto91}. In Table \ref{tab_nonxraydbrandom} we also
list one additional source (\#4359), which was eliminated from the random subsample because it was related
to the original target of the IPC observation (see
\S \ref{subsec_new}). This source is clearly extended in a {\it ROSAT} PSPC image and there are many faint
optical galaxies present on the Digitized Sky
Survey\footnote{\url{http://stdatu.stsci.edu/dss/dss\_form.html}} (DSS) at its sky location. However, a
few galaxies found in SIMBAD are at the same redshift as Abell 665, which was the target of the IPC
observation in which source \#4359 was discovered. Although this source is potentially scientifically
interesting (a galaxy group falling into Abell 665?), the EMSS procedure would have eliminated this source
and so we do as well. 
 
The random subsample source list was next correlated with X-ray source catalogs generated from the RASS
and other pointed observations made by {\it ROSAT} or {\it EXOSAT} using the facilities of the
HEASARC\footnote{\url{http://legacy. gsfc.nasa.gov/cgi-bin/W3Browse}} at NASA/Goddard Space Flight Center.
Many detections were made in the vicinity ($\leq$ 5 arcmin away) of random subsample sources, verifying
them as bona fide X-ray sources. However, none of these sources were clearly resolved and only two sources
(\#4845 \& \#3564) had hardness ratios that are comparable to hardness ratios of previous X-ray cluster
detections. Our optical imaging (described in \S \ref{subsec_opt}) of source
\#4845 finds no cluster present so that the hardness ratio is not considered further here as a definitive
discriminating characteristic. Table \ref{tab_xraydbrandom} presents the following information concerning
X-ray detections with S/N $> 4.0$ of the random sources made by other satellites (by column): (1) catalog
number of the source; (2) the X-ray satellite and instrument which made the detection; (3) the percentage
of the IPC flux that can be accounted for by the detected X-ray source(s) in the other instrument (in
making this calculation we have assumed  an energy index  of
$\alpha=1.0$, which is appropriate for an AGN; i.e., a hard cluster spectrum would contribute a larger
percentage, a soft stellar spectrum a smaller percentage than listed); the IPC flux used for the
calculation is taken from the smallest aperture which includes the {\it ROSAT} or {\it EXOSAT} detection;
(4) the angular offset between the random source centroid and the detected X-ray source (in arcmin); (5)
the 3$\sigma$ flux detection limit by the newer satellite image within 1.25 arcmin of the catalog source
(i.e., in the sky region of the first aperture) if there was no {\it ROSAT} detection made within that
region; (6) the database assessment of the angular extent of the detected X-ray source; (7) comments on
the detected X-ray source(s); and (8) the evaluation of the random source identification based upon these
X-ray detections alone. 

These new observations can be used to eliminate conclusively some sources from being clusters of galaxies.
Specifically, we eliminate all sources for which a reliable {\it ROSAT} detection (i.e., S/N $> 4.0$)  is
of sufficient flux to account for most or all of the IPC flux and is also time variable. Also PSPC
detections accounting for most or all of the source flux which are  found not to be extended sources are
unlikely clusters of galaxies, and have been eliminated.  This technique has been used successfully in
several serendipitous {\it ROSAT} cluster surveys
\citep[e.g.,][]{sch97a,ros98,vik98a}. However, there are some PSPC sources which may be marginally
extended and some that do not have determinations of extent available. These we have left unidentified.
All of the HRI sources in Table
\ref{tab_xraydbrandom} are listed in the {\it ROSAT} database as unresolved but, because the HRI is not
extremely sensitive to extended flux, we do not use this information as definitive unless the HRI source
accounts for the vast majority ($\geq 75\%$) of the IPC detection (i.e., there is little remaining flux
that could be extended). Additionally, we eliminate those sources as potential clusters for which new {\it
ROSAT} PSPC observations failed to detect a source within 3~arcmin of the catalogued source location to a
limit significantly less than the IPC detection in the smallest aperture. Either these sources are
variable and so are not clusters of galaxies, they are combinations of two or more sources flanking the
new X-ray source location, or they are spurious in some way. By these criteria, we have identified 17
sources in Table
\ref{tab_xraydbrandom} as not being clusters of galaxies (indicated with an X in column (8)).

\subsection{Optical Imaging of Sources in the Random Subsample\label{subsec_opt}}

To complete our study of the randomly selected sources, we conducted an optical imaging survey to
determine if any sources are in fact clusters of galaxies, the only plausible extended X-ray sources that
could have gone previously undetected (bright nearby galaxies could be extended X-ray sources but are well
catalogued and would have appeared in Table
\ref{tab_nonxraydbrandom}).  Our objective was to obtain deep imaging of a sufficient fraction of the
random subsample of new sources to make a statistical estimate of the number of X-ray clusters of galaxies
in the catalog. Our target list for the imaging campaign was the 42 sources within the random subsample in
the available RA range (0h-12h) and with declinations above -20 degrees.  We note that our imaging program
was designed to detect clusters out to $z\approx0.8$.  The optical imaging campaign is described in detail
in \citet*[Paper 2, hereafter]{P3}.

The results of our imaging program are presented in Table
\ref{tab_2.1mrandom}, which lists by column: (1) the catalog number of the random subsample source; (2)
comments for some fields; and (3) our evaluation of the field.  In the majority of fields, the optical
imaging alone, or in combination with the X-ray and other database searches, conclusively shows that no
distant cluster is present.   Three fields (\#1757, \#2036, and \#4057) that were eliminated from being
clusters based upon supplemental information unrelated to the imaging campaign (e.g., the existence of
X-ray point sources found in the database investigations) are discussed in Appendix
\ref{sec_appendix}. However, we did discover apparent galaxy over-densities in some fields, whose analysis
we discuss briefly here (details of our method are given in Paper 2). Galaxies were detected, and colors
and magnitudes were measured, with the galaxy photometry program PPP \citep{yee91}. A color-magnitude
diagram was constructed for each field, allowing an estimate of the redshift of any clusters or groups
present, identified by a cluster-red-sequence (CRS, which is compared to the galaxy color models of
\citealt{kod97}, and
\citealt{kod98}; our procedure is nearly identical to the method of \citealt{gla00}), as well as their
over-density relative to the field, given by the B$_{gc}$ statistic,
\citep{yee99}.  We present a discussion of the two sources we have identified as clusters or groups of
galaxies here to elucidate our identification criteria. Optical images, color-magnitude diagrams, and
detailed image analyses for these two sources are given in Paper 2. Unless otherwise stated, all fluxes
are in the \eband{} energy band, unabsorbed, (corrected for absorption assuming galactic neutral hydrogen
column density n$_H$ given by W3nH
\footnote{Neutral hydrogen data is from \cite{dic90}. W3nH is a Web version of the nH FTOOL. nH was
developed by Lorella Angelini at the HEASARC. It is a service of the Laboratory for High Energy
Astrophysics (LHEA) at NASA/GSFC and the High Energy Astrophysics Division of the Smithsonian
Astrophysical Observatory (SAO).
}). Luminosities are K-corrected assuming a power-law spectrum with photon index $\Gamma=1.5$
($\alpha=0.5$, following H92; consistent with a cluster spectrum in this bandpass) and quoted in the
\eband{} energy band in the rest frame of the cluster.

Source \#97: This source is identified as a nearby group of galaxies at $z=0.018$ (NGC 181/183/184). We
use the WPIMMS
\footnote{W3PIMMS is a Web version of the PIMMS (v3.0) tool. PIMMS was developed by Koji Mukai at the 
HEASARC. The first Web version was developed at the SAX Data Center. The SAX PIMMS package was ported to
and modified for the HEASARC Web site by Michael Arida. It is a service of LHEA.
}
software to convert the third aperture count rate for this source to an unabsorbed X-ray flux of f$_X=7.5
\times 10^{-13}$ \fx{} in the \eband{} band. We also calculate an X-ray luminosity of $1.0 \times 10^{42}$
\lxh{} in the \eband{} band, quite comparable to recent detections of similar, nearby,
elliptical-dominated groups of galaxies \citep{mul98}. The low redshift of this group makes the observed
field of view too small to calculate a robust value for B$_{gc}$; however, its optical appearance is quite
similar to other poor groups of galaxies detected by {\it ROSAT}
\citep{zab98}. Additionally, the count rates for this source ramp to a significantly higher level in the
larger apertures, as expected for a group subtending
$\sim10$~arcmin on the sky, rather than if this source were associated with just the largest elliptical in
the group. A Green Bank catalog radio source appears associated with the dominant E galaxy as well.
Therefore, we identify this source as a nearby group of galaxies.

Source \#161: This source exhibits a high overdensity of galaxies across the field; however there appears
to be more than one physical structure based on galaxy color and projected density. Accordingly, our
over-density estimates have been corrected (to lower values) in an attempt to avoid contributions from
galaxies at different redshifts. The galaxy over-density measurement is directly proportional to the
number of galaxies detected in the field less the predicted field galaxy counts in that sky area, so we
have segregated galaxies by color to assign them to each of the two concentrations, and corrected our
over-density estimates by the fraction of galaxies in each structure (additional details can be found in
Paper 2). We find that the most dominant sequence of galaxies lies within the redshift range
$z=0.52-0.59$. At a redshift of $z=0.55$, we measure a galaxy over-density of
B$_{gc}=1340\pm560$~\bggmph{} \citep[for a description of the uncertainty on the value of B$_{gc}$,
see][]{yee99}. Combining the observed correlations of galaxy number density ($N_{0.5}$) with X-ray
luminosity
\citep[L$_X$;][]{edg91} and B$_{gc}$ value with galaxy number density \citep[B$_{gc}=33
N_{0.5}$;][]{har01}, we can make a rough estimate of the expected luminosity of such a galaxy
over-density: L$_X$ (0.3-3.5) $= 3\times10^{45}$ \lx, corresponding to a total X-ray flux f$_X=2.3 \times
10^{-12}$ \fx. We convert the observed IPC count rates into fluxes to obtain (1.4, 2.7, \& 5.1) $\times
10^{-13}$ \fx{} for the ($2\farcm5$, $4\farcm7$, \&
$8\farcm4$) apertures, respectively.  The estimated flux from this cluster is higher than the actual IPC
aperture fluxes observed for this source so that this cluster is easily rich enough to produce the
observed X-ray emission. The next richest concentration in the field lies in the redshift range
$z=0.32-0.38$. At a redshift of $z=0.35$, we measure a galaxy over-density of B$_{gc}=740\pm240$~\bggmph.
This corresponds to a luminosity of L$_X = 4.6\times10^{44}$ \lx, and a total f$_X=8.9\times 10^{-13}$
\fx{} (i.e. 38\% of the flux expected from the distant cluster).  Thus, the more distant structure should
dominate the expected X-ray flux. However, the apparent high redshift cluster is also
$90$~arcsec from our X-ray centroid, which suggests additional sources or extended emission could
contribute to the X-ray detection. A marginal {\it ROSAT } HRI detection in a short exposure (7.8 ksecs) of
this source is positionally consistent (17 arcsec N) with the two brightest cluster galaxies (BCGs) in the
more distant concentration and has a flux consistent with being the sole contributor to source \#161. 
Although not obviously extended, this source is detected at a S/N of $<3.0$ and so is not a reliable
indicator of extent. In addition, there is an NVSS radio source consistent with one of the BCGs. Lastly,
there is a galaxy approximately halfway between the position of the cluster and the X-ray centroid in
Table \ref{tab_randomsample}. This object has
$r-i$ color consistent with being a member of the $z=0.55$ cluster, however its $g-r$ color is more than a
full magnitude bluer. This suggests it is an AGN, but it is neither X-ray bright (not appearing as a source
in the same exposure in which the BCG-centered source was detected) nor radio-loud (there is no NVSS
source at its location though one of the BCGs was detected). Thus we can only assume that it does not
contribute to the X-ray emission.  The combined data therefore suggest that the more distant galaxy
overdensity is of significantly higher flux than the nearer one, and is responsible for the IPC detection
in this field. Due to the difference in position of the apparent cluster and the X-ray centroid, and the
possibility of contamination by a second cluster or an AGN, we identify source
\#161 as a possible cluster of galaxies at $z=0.52-0.59$.

In summary, the random sample of 73 sources contains one possible cluster of galaxies (\#161, at
$z=0.52-0.59$) and one nearby group of galaxies (\#97, at $z=0.018$). The imaging survey, the database
search and the cross-correlation with existing {\it ROSAT} data eliminates a total of 41 of the 73 sources
from being clusters of galaxies. The remaining 30 have insufficient data at present to make an accurate
assessment of the presence or absence of a cluster of galaxies. These 30 are in no way different from the
rest, so that our sampling of this random subsample is itself random. Therefore, based upon our detailed
evaluation of a random subsample of these new catalog sources we find one possible cluster (\#161) and
thus 0--1 out of 43 ($\leq2.3\%$) sources are distant ($z\geq0.14$, the limit for inclusion in the EMSS
XLF evolution calculations) clusters of galaxies. 

\section{Investigation of Non-Random ``Selected'' Subsamples of New Sources\label{sec_nonrandom}} 

The small number statistics in \S \ref{sec_invest}  are consistent with zero, or up to 18 ($2.3\% \times
772$ sources) additional distant clusters among the 772 sources in this catalog. The addition of 18
possible clusters would represent a significant change to the cluster XLF at high redshift. Therefore, it
was necessary to hone our investigation to determine conclusively whether the catalog contains a
significant number of new, distant clusters or not. To this end, we investigated two non-random
subsamples, which were specifically chosen to maximize the selection of clusters and provide a robust
lower limit to the total number of cluster to be added to the XLF. In this section we describe the
investigation of sources within these two subsamples. 

\subsection{The ``Ramp'' Subsample\label{subsec_ramp}}

Assuming that a highly extended (or low surface brightness) source should increase its total flux (and
therefore S/N given a uniform background) with increased aperture size, we selected a sample of sources
wherein the S/N values increased (``ramped'') in larger apertures.  Out of some 461 sources with
appropriate RA and Dec. for our observing campaign, a total of 68 sources were found whose S/N in the four
IPC apertures ``ramped'' appropriately. That so few new sources exhibit the S/N behavior expected for
extended sources is another indication that, as found in the random sample investigation, a typical
catalog source is {\bf not} a cluster of galaxies. But even this ``ramping'' behavior could mean either
that flux from some point source offset from the catalog position was being detected in the larger
aperture (although the two point sources would have to be separated by the correct amount on the sky to
allow the S/N  to increase systematically for all three apertures) or that more and more flux from a
single, extended source was being included. It is also possible that other systematic effects could create
changes in the S/N in any aperture, either masking or creating apparently extended sources. 

Each of the 68 target ``ramp'' fields was investigated and given a ranking for observational priority. The
investigation for each field included inspection of the DSS image for that field; a search for any known
objects within 5 arcmin of the source location  within the NED and SIMBAD databases; and an accounting of
any known X-ray sources within the field from all public {\it ROSAT} and {\it Einstein} databases.  Fields
given first observational priority could have included some evidence of a galaxy over-density in their DSS
image, and no evidence for multiple X-ray point sources in the field (which would indicate a combination
of X-ray sources responsible for the new IPC detection). On this basis, 17 sources were assigned the
highest priority, 6 of which were successfully observed. Lowest priority was assigned to fields with known
QSOs or other non-cluster X-ray emitters present; none of these fields were observed. The remaining
sources, lacking either of the above indicators, were placed as second priority; and 8 of these sources
were also observed. The ``ramp'' sources observed are all typical of the first and second priorities, with
individual fields chosen for observation simply on the basis of their sky availability during LST ranges
not heavily populated with random sources listed in Table
\ref{tab_randomsample}. The basic catalog X-ray data for the 14 sources observed optically (less than one
quarter of the full ``ramp'' subsample), are listed in Table
\ref{tab_rampsample}, with the same columns as Table
\ref{tab_randomsample}. We note that there was no attempt in the creation of the ``ramp'' source list or
in the observation of these sources to favor sources with high flux; i.e., the ``ramp'' sources  are
typical of catalog median values in their aperture count rates.  

A summary of the database search and optical imaging results for the 14 observed fields are shown in Table
\ref{tab_2.1mramp} with the same columns as in Table
\ref{tab_2.1mrandom}. For each cluster candidate, the X-ray luminosity in the IPC bandpass of
$0.3-3.5$~keV was estimated from both the optical galaxy over-density B$_{gc}$, and the measured source
count-rate in the IPC apertures (under the same assumptions used for source \#161 described above). To be
identified as a cluster of galaxies, the galaxy over-density was required to account for the majority of
the observed X-ray flux. Four target fields (\#992, \#1310, \#1492, and \#1605) had apparent galaxy
over-densities, and were the only ``ramp'' sources to be identified as clusters of galaxies. Images,
color-magnitude diagrams, and details of the image analysis for individual fields are given in Paper 2.
Note that source \#992 was found to be target related, and was eliminated from the catalog (details are
given in Appendix
\ref{sec_appendix}).

With four distant clusters found in the examination of only 14 ``ramp'' sources, or
$\sim$30\% of candidates investigated, we find that this an extremely viable cluster selection technique.

\subsection{The ``High Flux/Signal-to-Noise'' Subsample\label{subsec_hiflux}}

Based upon the confusion statistics presented in \S~\ref{subsec_est} and Table
\ref{tab_confusion} as well as the random subsample investigation presented in
\S~\ref{sec_invest}, a typical catalog source is likely to be either a combination of two or more lower
flux sources and/or a single point source, whose presence in the new source catalog, but not in the EMSS,
is due to the vagaries of small number statistics.  With these two concerns in mind, we created a
subsample of the catalog consisting of those sources with either the 50 highest X-ray fluxes, and/or the
50 highest S/N values. There was only modest overlap between the two criteria, resulting in a subsample of
92 sources. The highest S/N sources are the most likely sources not to be affected by small number
statistics and so least likely to be single point sources not in the EMSS but present here (see
\S~\ref{subsec_moreacc} for more details on this important point). The highest flux sources  are those
least affected by confusion (see Table~\ref{tab_confusion}). This source list comprises the best
candidates for the purpose of discovering previously unknown rich clusters of galaxies in this catalog. We
suggest that subsequent investigations of this catalog concentrate on these sources. We list in Table
\ref{tab_highsample} the basic X-ray data for those sources which we observed from this sample. Columns
are the same as in Table
\ref{tab_rampsample}.  We also note that two of the ``Ramp'' sources (\#1492 and \#1605) are also part of
the high flux or signal-to-noise ratio (HFS) subsample, and that they were observed optically as a part of
the ``Ramp'' subsample.

An optical imaging program for the HFS subsample, using similar priorities to those described in
\S~\ref{subsec_ramp} was undertaken in good weather conditions in May 2000. From a total of 20 observed
fields, 11 contained apparent galaxy overdensities, and were further scrutinized to identify potential
clusters. The results of the imaging program are shown in Table
\ref{tab_2.1mhigh}  (columns are identical to Tables
\ref{tab_2.1mrandom} and \ref{tab_2.1mramp}). X-ray luminosities for cluster candidates were calculated in
the same manner as for sources in the ``ramp'' subsample, and are given in Table \ref{tab_newclusters},
described below.  Three sources are identified as clusters or groups of galaxies, and two additional
sources are possible clusters due to some uncertainties within or inconsistencies among the observational
data. The remaining six cluster candidates were rejected after detailed investigation, and are discussed
in Appendix~\ref{sec_appendix}. Images, color-magnitude diagrams, and details of the image analysis for
individual fields identified as clusters are presented in Paper 2.

A database and literature search was conducted for all sources in the
(HFS) subsample and some clusters were found that had been previously identified. These sources are
discussed in Appendix \ref{sec_appendix}, and included in Table \ref{tab_newclusters} below. In addition,
3 of the 11 apparent galaxy overdensities described above had a corresponding RASS Bright Source Catalog
(BSC) or Faint Source Catalog (FSC) detection. These detections are listed in Table \ref{tab_2.1mhigh}.

\subsection{Identifications from Other X-ray Surveys\label{subsec_ident_other_surveys}}

In addition to the systematic investigations of pre-selected subsamples, we have also checked our source
list versus on-going cluster identification programs by 
\citet[The Wide Angle {\it ROSAT} Pointed Survey I (WARPS I)]{per00},
\citeauthor{rom00} (\citeyear{rom00}; the Serendipitous High-Redshift Archival {\it ROSAT} Cluster 
(SHARC) group\footnote{\url{http://www.astro.nwu.edu/sharc/}}), and the 160 Square Degree {\it ROSAT}
survey \citep[160SDS, hereafter,][]{vik98a,vik98b}.  Each of these programs list clusters identified from
serendipitously detected {\it ROSAT} PSPC sources.  There were six sources found in both our catalog and
that of another group (\#420,
\#2203, \#2616, \#2626, \#3175, and \#3469), each is discussed in Appendix
\ref{sec_appendix}. Sources \#420, \#2203, \#2626, and \#3469 are identified as clusters of galaxies, and
we adopt their measured redshifts and luminosities (where available).

\begin{\myfigure}
\centerline{\epsfig{file=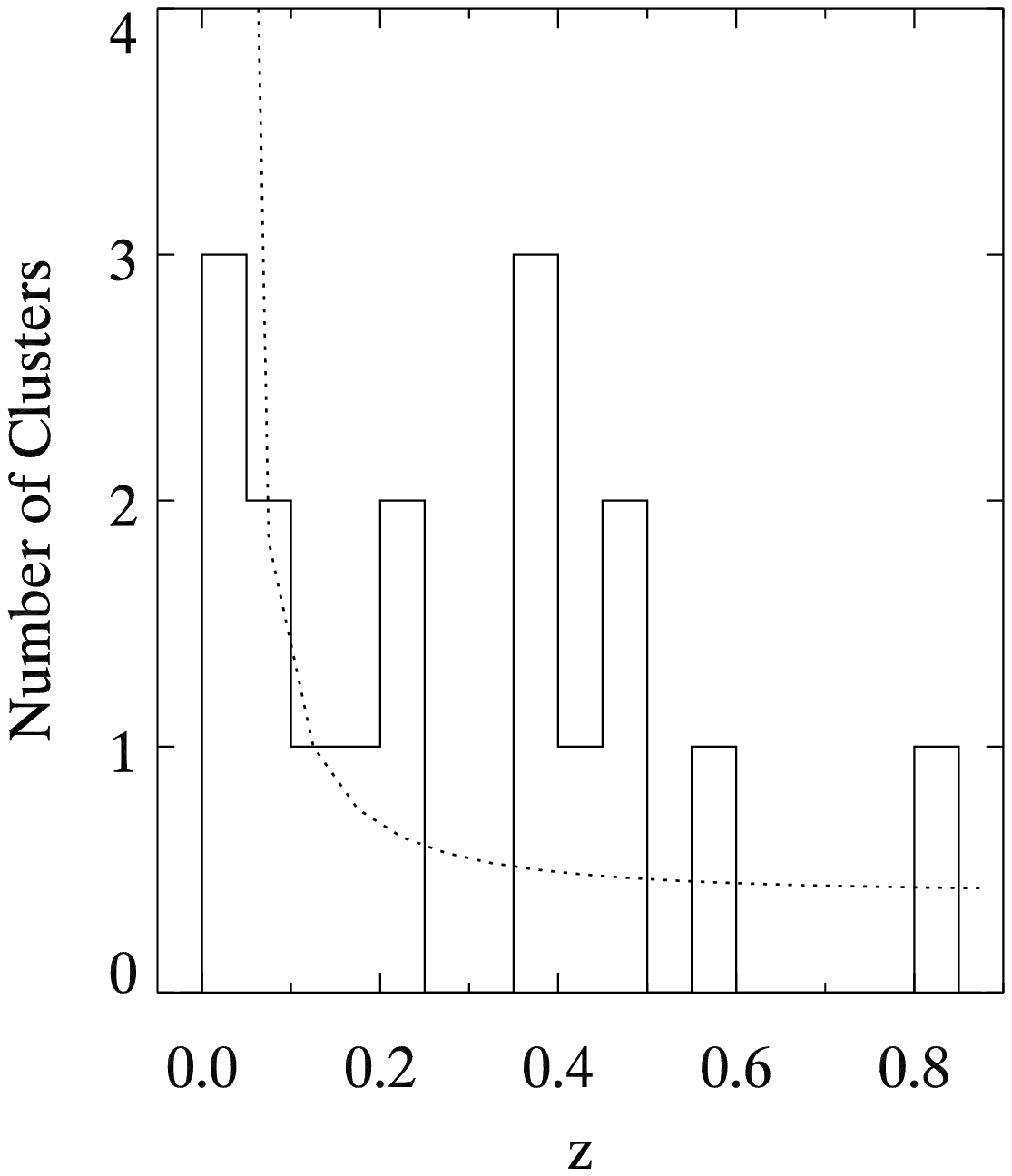, scale=0.50}}
\caption[The Redshift Distribution of Newly-Discovered EMSS Clusters]{The redshift distribution of
newly-discovered EMSS clusters; data from Table
\ref{tab_newclusters}. The dashed line overlaid on the histogram indicates the expected number of missing
clusters based on the ratio of EMSS detect cell flux to total flux (see \S \ref{subsec_sumnonrandom}),
normalized to the number of objects in the bin containing $z=0.14$. The distant, luminous
clusters found in this work are in clear excess over that expected based on the H92 methodology in the
redshift range ($0.3\leq z\leq0.6$).}
\label{fig_zhistogram}
\end{\myfigure}

\epsscale{1.0}
\begin{\myfigure}
\centerline{\epsfig{file=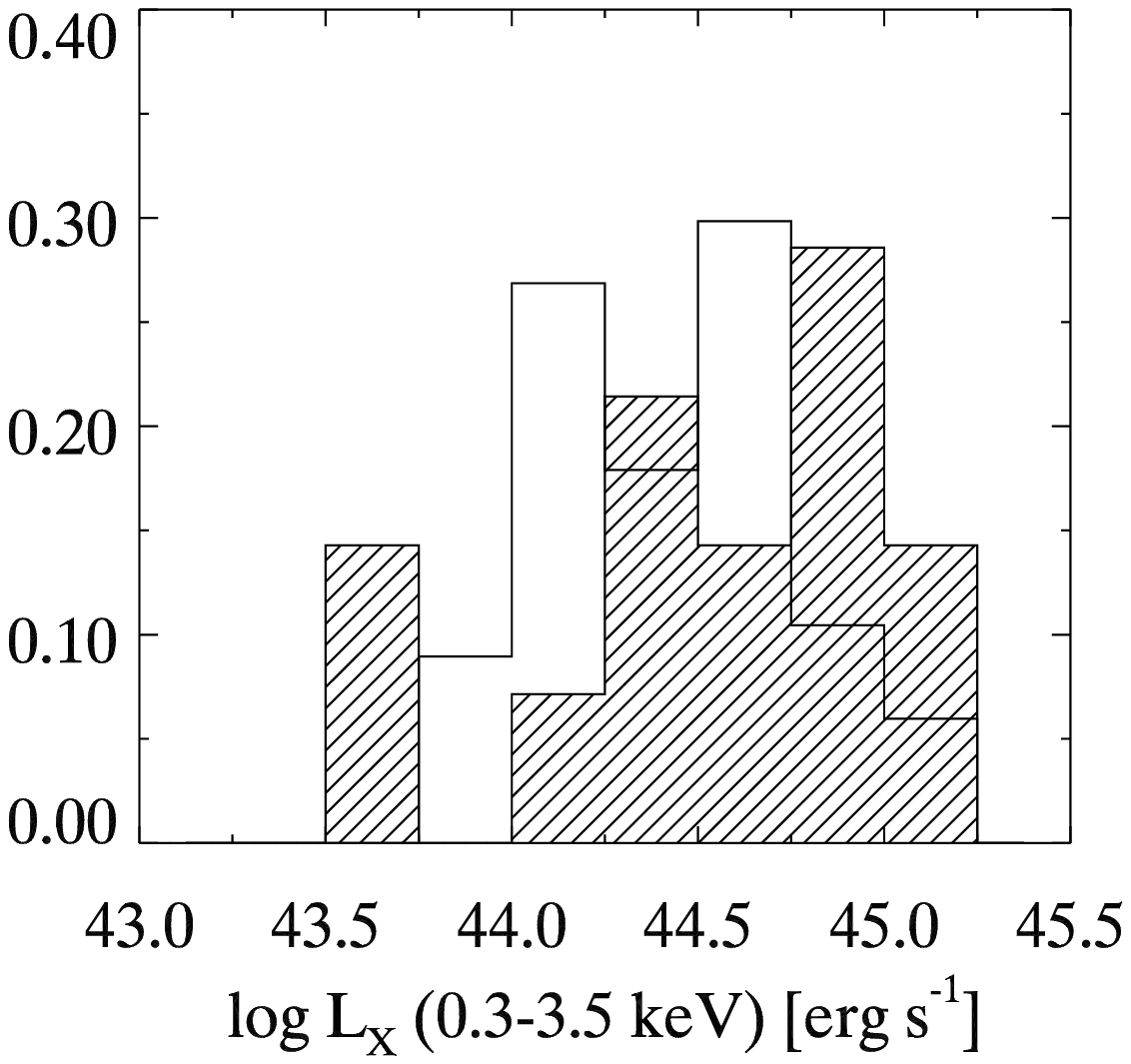, scale=0.50}}
\caption[The Luminosity Distribution of Newly-Discovered EMSS Clusters]{The normalized luminosity
distribution of newly-discovered EMSS clusters (shaded histogram); data from Table
\ref{tab_newclusters}. Overlaid is the normalized luminosity distribution of the original (H92) distant
($z\geq0.14$) EMSS clusters (open histogram). The new clusters have very similar X-ray luminosities to the
original EMSS clusters.}
\label{fig_lhistogram}
\end{\myfigure}
\epsscale{1.0}

\subsection{Summary of Non-Random Subsamples\label{subsec_sumnonrandom}}

In summary, our optical observations of pre-selected subsamples and investigation of available databases
have found eight new or previously known clusters at $z\geq0.14$, two
$z\geq0.14$ possible clusters, and five
$z<0.14$ clusters or groups (this accounting excludes Source \#992, which was eliminated from the catalog,
see Appendix \ref{sec_appendix}). Six of these sources are confirmed X-ray sources by {\it ROSAT}
observations. A summary of data for these clusters (as well as those found in the random subsample) is
shown in Table
\ref{tab_newclusters} which includes by column: (1) catalog source number; (2) spectroscopic redshift or
estimated redshift range from photometry; (3) measured galaxy over-density, B$_{gc}$ in
\bggmph; (4) log of the X-ray luminosity in \lxh{} in the
\eband{} bandpass calculated from the third IPC aperture flux; (5) log of the X-ray luminosity in \lx{} in
the \eband{} bandpass either estimated from the B$_{gc}$ value (see \S~\ref{subsec_opt}), or as given by
the reference for the source; (6) the subsample from which the cluster was identified; (7) notes on the
nature of or identification for the cluster; and (8) the reference for the cluster discovery. We have used
the data from the third IPC aperture to determine a total X-ray luminosity for these distant clusters.
From the previously detected EMSS clusters at $z\gtrsim0.2$
\citep*{P4}, we find core radii of $\leq50$ arcsec so that $\gtrsim90\%$ of the X-ray flux is contained
easily within the third aperture assuming a standard $\beta$ model. Even though some of these clusters may
be more diffuse than previously detected EMSS clusters, the third aperture should contain the great
majority of their flux.

For each cluster discovered in our optical imaging program, the detailed justification of the ID,
estimated redshift, galaxy over-density, and X-ray luminosity is presented in Paper 2. Fields
with apparent galaxy over-densities which nonetheless we identified as not being clusters are discussed
in Appendix \ref{sec_appendix}.  While this investigation is far from a complete accounting (i.e., only
approximately 30\% of the ``ramp'' and HFS subsample sources have been imaged, and the HFS subsample has
an arbitrary lower limit), it is clear that this catalog does contain some previously undiscovered rich
clusters. In Figure
\ref{fig_zhistogram} we present the redshift distribution of the clusters in Table
\ref{tab_newclusters}. 

Based on the ratio of a cluster's measured flux in the $2.4 \times 2.4$ arcmin EMSS
detect cell ($F_{det}$) to its total extrapolated flux ($F_{tot}$), we can estimate the expectation for
the EMSS to miss extended objects. The calculation of $F_{tot}/F_{det}$ is shown in Figure 1 of H92 as a
function of redshift. If we normalize this function to the number of clusters we have now discovered at
redshift $z\approx0.14$, it should indicate the relative amount of flux not measured, and therefore
provide an estimate of the relative number of clusters which would be missed at other redshifts by the
EMSS detect cell {\it if all clusters have surface brightness profiles described by a $\beta$-model with
$\beta=0.67$, and $r_{core}=250$~\hkpc,} the canonical values assumed by H92. We have overlaid the curve
of $F_{tot}/F_{det}$ on Figure
\ref{fig_zhistogram} (the dashed curve), showing the expected number of missing clusters, normalizing the
curve to the histogram bin containing
$z=0.14$. Figure \ref{fig_zhistogram} shows the surprising result that the majority of new clusters are at
a redshift
$z\gtrsim0.35$, well beyond the $z\sim0.1$ regime where one would expect to have missed clusters
by using the formalism of H92. This result is unexpected only if clusters at all redshifts have similar
structure (i.e., similar core radii and
$\beta$ values).
The presence of so many new clusters at
$z\geq0.35$ may be due to the high$-z$ clusters being more diffuse, and thereby missed in the EMSS due to
a selection bias against low surface brightness sources. In detailed simulations, \citet{ada00} have shown
that low surface brightness clusters (e.g., clusters with $\beta = 0.55$ or $r_{core} = 400
h_{50}^{-1}$~kpc) have a significantly lower detection efficiency using the SHARC detection method, which
is a wavelet-based algorithm specifically designed to detect extended objects. We expect that the EMSS
method has an even more pronounced loss of efficiency for such clusters. Even if the new clusters are not
extremely diffuse (e.g., \citealt{vik98b}, using an algorithm designed to detect extended objects,
detected no significant increase in measured cluster core radius, assuming a standard
$\beta$-model with $\beta=2/3$ for clusters at redshifts $z>0.4$ in the 160SDS compared with
the nearby luminous sample of
\citealt{jon99}), it is plausible that the selection bias against extended sources was more severe than
anticipated by the EMSS team. Due to the existence of high-$z$ clusters near the flux limits of the IPC
observations, the EMSS survey missed several clusters which we are now discovering. In Figure
\ref{fig_lhistogram} we present the normalized luminosity distribution of the clusters in Table
\ref{tab_newclusters} (shaded histogram). We have also shown the normalized distribution in luminosity of
distant ($z\geq0.14$) EMSS clusters taken from H92 (open histogram). We can see from Figure
\ref{fig_lhistogram} that the new clusters exhibit a similar range of luminosities as the original distant
EMSS clusters.

\section{Implications for the EMSS Cluster Sample\label{sec_impl}}

In this section we evaluate the results of this incomplete investigation of new X-ray sources in EMSS
fields. We use the results above to make the most accurate determination of the actual number of new
sources that could potentially be added to the EMSS. That is, based upon our detailed study of the random
and non-random samples, some of the sources in the catalog would not have survived the detailed scrutiny
to which EMSS sources were subjected (see Appendix
\ref{sec_appendix} for individual examples). These must be eliminated, at least statistically, for
comparison with the EMSS. We also use the number and estimated redshifts and luminosities of the new
distant clusters we have found to infer the statistical impact on the EMSS XLF and its evolution.

\subsection{A More Accurate Comparison Sample for the EMSS Source Catalog
\label{subsec_moreacc}}

Based upon the investigation of the catalog subsamples described in the previous sections, it appears
that: (1) many of the catalog entries are the ``confused'' combinations of two or more fainter sources 
and thus not true additions to the EMSS; (2) many sources are single, likely point-like sources, which
were not included in the EMSS for reasons as yet unclear (but perhaps due only to Poisson statistical
fluctuations occurring between different size and shape apertures);  (3) $\leq2.3\%$ of the new catalog
sources may possibly be distant clusters of galaxies, and we have found several examples. A similar number
of nearby ($z<0.14$) poor clusters or groups of galaxies are likely to be present. If this is the case, and
considering that the source catalog was generated from only a subset of the EMSS sky area, the impact on
the EMSS X-ray luminosity function (XLF) could be significant. In order to estimate the impact on the EMSS
cluster XLF as precisely as possible, in this section we scrutinize the catalog in more detail based upon
our investigations in \S~\ref{sec_invest} \& \ref{sec_nonrandom}.  We will attempt to emulate the  EMSS
selection criteria as precisely as possible using knowledge gained from problems found in the random and
non-random subsamples. This process will help us estimate the most accurate number of new sources to be
added to the EMSS fields. 

However, we emphasize that almost all of the corrections in this section are statistical in nature; i.e.,
it is not possible to determine which individual sources should be removed from the catalog (and which
should stay) in order to make the most accurate comparison -- we can only estimate the fraction of sources
which should be removed.  Therefore, the results of this section do not discredit the reliability of any
individual source in the catalog; i.e., based  upon the catalog evaluation presented in
\S~\ref{sec_invest}, we believe that the large majority of these sources are real fluctuations above
background, although most are superpositions of fainter sources.

We have edited the catalog (originally consisting of 772 sources) into a final EMSS comparison sample of
406 sources by the following four actions:

1. While we have shown in \S~\ref{subsec_new} that the methods employed herein do not significantly bias
the detected source counts  relative to the EMSS, they have in fact modified the source detection method
and background determination such that the S/N is higher for sources detected using the OHG97 algorithms
when compared with EMSS values. 
Thus the catalog contains
new sources within the EMSS sky area simply because faint X-ray sources in the same fields are now
detected at
$\geq4\sigma$ while the EMSS detected them at $<4\sigma$. This is the most likely reason why our imaging
survey has concluded that some sources are single, point-like X-ray emitters. Since these sources appear
in the new catalog only due to more favorable  statistics, we conclude that these sources should not be
added to the EMSS.  Because it is impossible to identify exactly which sources these are, and in order to
account for this bias relative to the EMSS, we artificially raise the detection limit for the larger
apertures to correct for this effect statistically.  To determine the new detection limit for the larger
apertures, we show in Figure~\ref{fig_3sigmaplot} the S/N of 334 EMSS sources thought to be point-like on
the basis of their optical identifications (the same sample used for Figure \ref{fig_cr}, see \S
\ref{subsec_new}). Figure \ref{fig_3sigmaplot} plots the original EMSS S/N versus the  S/N of these same
sources measured in the three smallest (2.5, 4.7 \& 8.4 arcmin in diameter) IPC apertures. Since these
EMSS sources are identified as single, unresolved X-ray sources, the differences found should only be due
to the different detect cell sizes and detections methods employed herein. A similar plot, but for all
redetected EMSS sources, can be found as Figure 5 in OHG97.

Figure \ref{fig_3sigmaplot} presents a good correlation between the signal-to-noise ratios determined by
these different techniques, which reveals a mild systematic bias in favor of the new background
determinations and the circular apertures used herein (i.e., a systematically higher S/N compared to the
original EMSS) but with a substantial scatter as well. Evidently, the Poisson statistical regime plus the
slight differences in detect and background cell sizes, shapes, and locations  and the different
background maps combine so that some EMSS sources are redetected at higher S/N by the current technique
and some are redetected at slightly lower S/N. The possibility of such variation in measured S/N was
raised in \S~\ref{subsec_random} where some random subsample sources were found to have a S/N $\geq$ 4 in
the original EMSS detection aperture, although of course they were not detected by the EMSS. This
variation in measurement creates a dilemma for an accurate comparison between the present work and the
EMSS since we cannot say absolutely whether any one source close to the 4.0$\sigma$  detection limit
should be included within the EMSS or not. Instead of attempting to scrutinize each source individually to
estimate the EMSS S/N, we treat this bias statistically using the median ratios of S/N in the catalog
relative to the original EMSS values, shown as dotted lines in Fig.
\ref{fig_3sigmaplot}. The new limits which correspond to 4.0$\sigma$ in the EMSS are 4.1, 4.3 and
4.5$\sigma$ for the 2.5, 4.7, and 8.4 arcmin apertures, respectively.  We emphasize that these new
detection limits still do not emulate the EMSS as closely as one would like because of the significant
spread in S/N for the source detections, which are not related to the source being resolved with respect
to the detect cell. Thus, ``new sources'' would be detected in the EMSS sky area even if all sources were
point-like with respect to the detection aperture employed (and some ``old'' EMSS sources would not be
re-detected for the same reason).  Nevertheless, these new S/N limits statistically eliminate 261$\pm$16
sources from the comparison sample (including two sources identified as clusters of galaxies, \#1310 and
\#1492, and one low redshift galaxy group, \#97).

We also note that even though the elimination of these low-$\sigma$ sources provides a more accurate
comparison with the EMSS, these sources are nevertheless likely to be real because the background
determinations for them have been improved; they are just a bit too faint for secure inclusion in the EMSS.
 
2. As previously derived from theoretical considerations and verified in the imaging survey, many aperture
3 \& 4 detections are likely to be confused.  Some catalog sources were detected at $\geq$ 2.5$\sigma$ in
only apertures 3 and 4, which strongly suggests a combination of X-ray sources rather than a single
extended source. We note that 9 such sources are present in the random subsample, four of which have {\it
ROSAT} PSPC detections far enough away from the source position that the PSPC sources would not contribute
any flux to the smallest aperture. A fifth source has a 9th magnitude star $4\farcm6$ distant, which is a
very likely ID and again would not contribute any first aperture flux. None of the other 4 random sample
sources in this category had either optical or X-ray database information. Our interpretation is that
these catalogued sources are a confusion of a real (but $<4\sigma$) offset source with other emission.
What should have been a single source below the detection limit was buoyed up by  circumstance, (e.g.,
proximity of rib structures or catalog boundaries to the background apertures, or nearby faint discreet
sources) thus increasing the S/N in larger apertures above $4\sigma$.  Had the offset source been
$\geq4\sigma$ by itself, the source centroid should have been closer to it. The automated detection and
centering algorithms are so accurate that there are very few of this type of source. Therefore, we
eliminate all such sources (53 in number) from the comparison sample.

3. The issue of overlapping (but not coincident, or merged) observations used to generate this catalog
which were not included in the original EMSS was brought up in \S \ref{subsec_new} but not addressed for
the entire catalog. Though it is not feasible to investigate all such occurrences, we have instead
conducted a detailed scrutiny of individual sources in the random subsample to identify how many sources
suffer from such an overlap. We then estimate the decrease in S/N for each such source that would result
from eliminating the non-EMSS observations. From the 73 sources in the random subsample, there were a
total of 11 sources with overlapping fields. Nine sources were overlapped with rib-free regions of
non-EMSS fields, and thus have potential extra flux from those exposures. There was no indication of
variability in any of these 9 sources based on inspection of each sub-exposure.  The remaining 2 sources
had rib-edge code difficulties in one or more of the observations.

Of the nine sources with normal overlapping: six had exposure time corrections that would reduce the S/N
in the first three apertures  to $< 4.0$, thereby eliminating them from the catalog; two had exposure time
corrections that would not change their status -- they remain in the catalog; the two remaining sources
were overlapped by a rib-region of the IPC field, resulting in vignetting. One of these sources would
remain a catalog source, assuming a maximum correction to the S/N based on the total exposure time. The
other would be removed ($<4.0\sigma$) under the same assumptions, although the actual vignetting is not
known and would lessen the affect of this estimate. Any undetected source variability could affect the
estimates significantly by increasing or decreasing the amount of flux detected during valid EMSS
observations relative to the total flux.  Thus we estimate that
$10\pm1.4\%$ ($7\pm1$ of 73) of the random subsample catalog are present due to the addition of non-EMSS
overlapping exposures, and should therefore be eliminated from the comparison sample.

Because it was not possible to scrutinize the entire source catalog at the same level of attention as the
random subsample, we have assumed that a similar percentage of sources in the final catalog suffer from
the overlap scenario (i.e., $10\pm1.4\%$). Therefore, although we cannot identify which individual sources
these are, we reduce the number of catalog sources for statistical purposes by 10\% to account for these
additional problems.

4. There was one source listed in Table \ref{tab_randomsample} (\#4359) which, after significant optical
followup and database scrutiny, was eliminated as being target related. In this case, the source appears
to be a portion of the X-ray emission from Abell 665, the target of the original IPC observation (see
Appendix~\ref{sec_appendix}). Another two such sources were found in the detailed investigation of
non-random sources, and so this effect could occur in modest numbers in the full catalog. A few similar
sources were removed from the EMSS after some optical and database followup work and could not have been
found prior to such detailed investigation. Since all catalog sources have not been scrutinized in this
way, only a statistical culling is possible. On the basis of the random subsample evaluation, we eliminate
1.4\% (1 of 73) of sources as being target related in this pernicious manner. 

After culling the catalog using these new criteria, the number of sources in the final comparison
sample is 406$\pm$17.  The error estimate on this number is an in-quadrature combination of the
various error estimates for the individual statistical subtractions made above. This number is
comparable to the 478 EMSS sources found in the same sky area. We emphasize that this number is
statistical in nature and (except for point \#2 above) we cannot with high confidence identify
individual sources which should be eliminated from the catalog. Therefore, we retain the full source
catalog and will make it available electronically upon request to the first author.

\begin{\myfigure}
\centerline{\epsfig{file=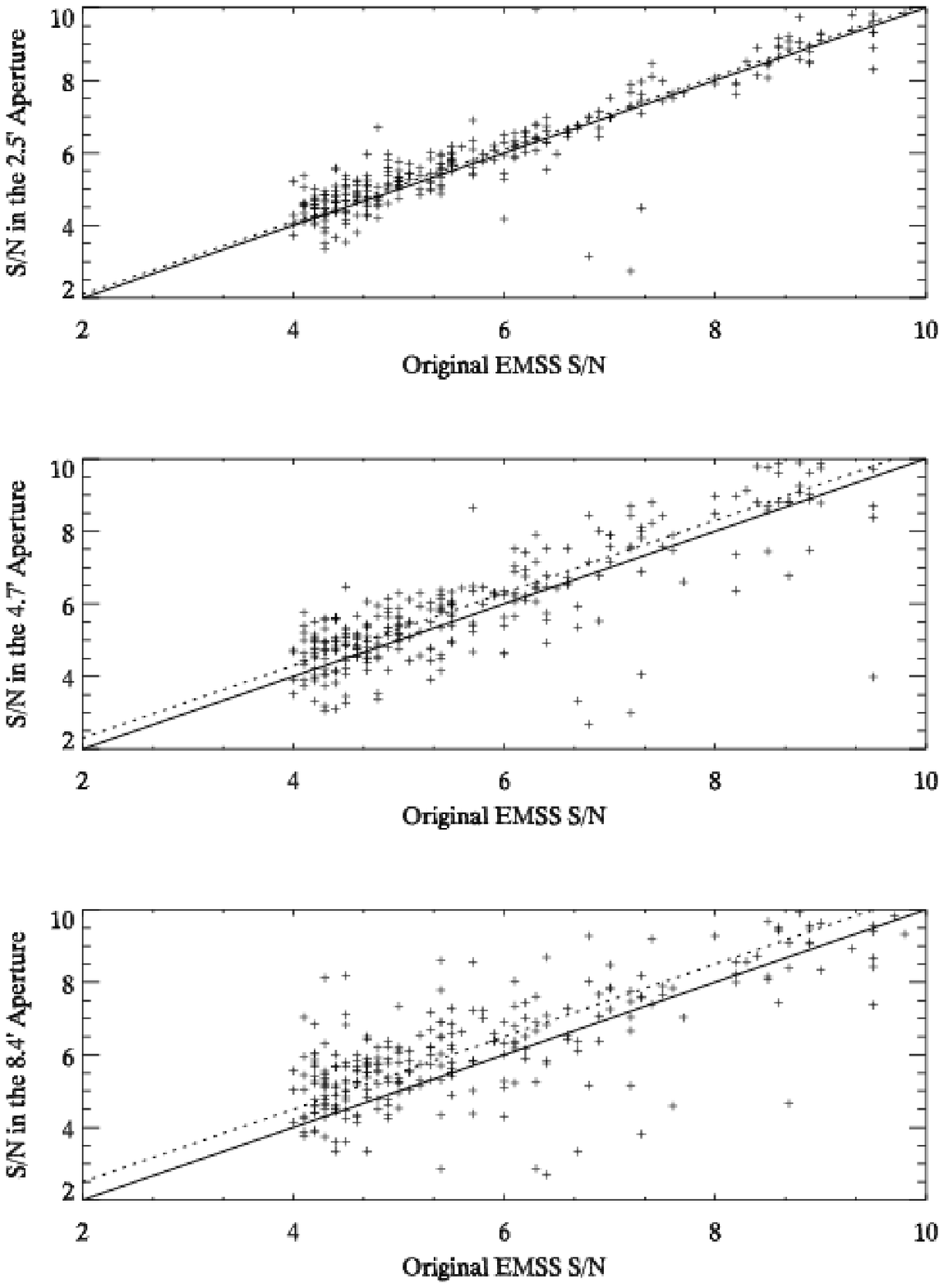, scale=0.65}}
\caption[S/N Comparison of point-like EMSS sources]{S/N of 334 point-like EMSS sources tabulated in the
EMSS catalog vs. S/N measured in the 3 smallest apertures used in this work. The solid line is the
one-to-one ratio expected if there were no bias between apertures. The median bias found in each aperture
is shown by a dotted line; relative to the EMSS detection limit of 4.0$\sigma$, the bias corresponds to
values of 4.1, 4.3, and
$4.5\sigma$ for the 2.5, 4.7, and 8.4 arcmin apertures, respectively. However, note the significant spread
in ratio values for all apertures.
\label{fig_3sigmaplot}}
\end{\myfigure}
\epsscale{1.0}
\smallskip
\smallskip

\subsection{The Effects on the EMSS X-ray Luminosity Function (XLF)\label{subsec_xlf}}

Applying the $\leq2.3\%$ estimate of distant clusters found among the random sampling of new sources in \S
\ref{subsec_opt} to the final number of EMSS comparison sources (406) arrived at in \S
\ref{subsec_moreacc} suggests that there are a total of $\leq9.4$ ``new'' clusters in the comparison
sample for the EMSS (and $\leq18.0$ clusters in the entire 772 entry catalog). A comparable number of
low-redshift ($z<0.14$) poor clusters and groups are also expected based upon our random sampling. Our
investigations have already found 11 new distant clusters and 6 nearby clusters or groups in the entire
catalog. Counting the three possible clusters as only a one-half detection, and removing objects which
would not pass the S/N cut made in \S~\ref{subsec_moreacc} (\#97, \#1310, \& \#1492), reduces these
numbers to 7.5 and 5, respectively. All of  our distant cluster detections above the S/N cut are in the
HFS subsample  excepting the possible cluster \#161.

Thus we may set a value of 7.5 new distant clusters as the firm minimum addition to the EMSS sample.
However, we have only partly investigated our catalog. Assuming that new cluster detections will only come
from the HFS subsample (which is conservative; source \#161 may be a distant cluster, and is not part of
the HFS subsample), we can estimate how many clusters are in the full HFS subsample based upon how much of
that sample we have yet to observe. Our optical observations to date have allowed evaluations of 26 fields
in the HFS subsample (although Table
\ref{tab_2.1mhigh} has only 20 entries, 6 sources observed as part of our investigations of the ``Ramp''
and Random subsamples are also HFS sources). In addition our database and literature investigations have
resulted in 15 further identifications. Thus 51 of 92 HFS sources remain unidentified and unobserved
optically.  Assuming the same distribution of identifications as the observed sample, we estimate that
within the remainder of the HFS subsample 5.6 distant clusters, 
1.9 nearby clusters or groups,
3.7 possible clusters,
and 31.8 non-clusters are present.
%
Reducing these numbers by $10\pm1.4\%$ to account for sources in overlapping fields (see
\S~\ref{subsec_moreacc}) results in 5 distant clusters, 1.7 nearby clusters or groups, 3.3
possible clusters, and 28.6 non-clusters. Counting possible clusters with half weight, we arrive at a
range of $7.5-14.2$ new clusters in 
the entire catalog. This number is clearly conservative, assuming as it does that all new clusters will
only be found in the HFS subsample, with no clusters below the arbitrary flux/count rate limit of that
sample. Since the EMSS detected 37 distant and 12 nearby clusters in the same sky area as surveyed by this
reanalysis, the EMSS distant cluster sample as presented in H92 is estimated to be 72--83\% complete. 

The nearby poor clusters or groups of galaxies we have discovered were missed by the original EMSS because
they are so large on the sky that most of their flux falls outside of the EMSS detect cell. This
eventuality was foreseen by \citet{gio90b} and H92 who excluded the redshift range $z\leq 0.14$ from the
EMSS XLF determination because of the very large  correction that was needed to correct the EMSS detect
cell flux to a total flux. In addition, the original EMSS catalog did not expect to be complete at $z\leq
0.14$ due to the fact that clusters of galaxies at these redshifts were frequently the targets of the IPC
observations and therefore excluded from the serendipitous EMSS catalog. Thus the new
$z\leq0.14$ clusters found in the catalog will not impact the EMSS XLF determination at all. However, the
addition of new, distant clusters to the EMSS could modify the XLF significantly, particularly in the
high$-z$, high$-$L$_X$ bins, where the current EMSS cluster numbers are quite small. 

In order to construct a revised cluster XLF, we must first update the EMSS XLF, based upon work more
recent than H92. The H92 XLF sample of clusters contained 67 clusters at
$z>0.14$, and L$_X>5\times10^{43}$ \lxh. The sample was binned for the calculation of the luminosity
function at different redshifts. Three redshifts bins at ($0.14<z<0.20$), ($0.20<z<0.30$), and
($0.30<z<0.60$) were each subdivided into 6 luminosity bins, ($0.5-1.0$), ($1.0-2.0$), ($2.0-3.98$),
($3.98-7.94$), ($7.94-15.85$), and ($15.85-31.62$)$\times10^{44}$ \lx.  Since then, subsequent work has
modified these original cluster data somewhat; details of the changes and updates can be found in
\citet{thesis}. These changes include new or revised cluster redshifts and/or luminosities measured since
H92, as well as revised identifications primarily based upon a systematic {\it ROSAT} HRI imaging program
of possible BL Lac objects originally identified as clusters of galaxies \citep{rec99}. One source,
MS~1317.0-2111, was not detected in the {\it ROSAT} HRI campaign for unknown reasons, and so we leave its
ID = cluster as in H92. In Figure \ref{fig_xlf_comp_update_h92} we show the EMSS XLF in three redshift
shells. Open circles are the original EMSS data taken directly from Figures 2 and 3 in H92, open triangles
reflect all of the corrections we have just described. We have recalculated the luminosity function using
the updated data, following in every detail the prescription for calculating search volumes as described in
\citet{gio90b}, H92, and \citet{hen00} in order to match the H92 results.  However, there are now clusters
in the EMSS sample at $z>0.60$, and so a fourth redshift shell ($0.60<z<0.85$) has been added to
accommodate all clusters now in the sample. We will truncate the volumes for clusters in the third shell
($0.30<z<0.60$) in exactly the same manner as H92. The resulting XLFs are shown in Figure
\ref{fig_xlf_comp_update_h92}. None of the values change beyond the 1~$\sigma$ errors, but note that the
largest changes are in the lowest luminosity bins.

After making these corrections, we now use the newly discovered clusters in this catalog to estimate the
full number of new EMSS clusters that should be added to each ($z$,$\,$L$_X$) bin in the full EMSS sky
area. The estimated redshift range of each distant cluster in Table
\ref{tab_newclusters} place it in a single redshift bin; however, our luminosity estimates in a few cases
span more than one luminosity bin. We have therefore added `fractional' clusters to the appropriate
luminosity bins. Furthermore, three of the clusters in Table
\ref{tab_newclusters} (\#161, \#2844, and \#2906) are only possible cluster identifications.
Conservatively assuming a 50\% false identification rate for possible clusters, we add one-half cluster to
the relevant ($z$,$\,$L$_X$) bins for each of these sources. 

Based on the results from \S~\ref{sec_impl}, there are
$\leq9.4$ distant clusters in the comparison sample of $406\pm17$ objects. Based on the actual number of
detected clusters in the catalog, and the estimate of clusters not yet found within the HFS subsample
given above, there are $7.5-14.2$ distant clusters in the catalog.  As the two estimates are consistent,
we will adopt the latter as it is based on a larger number of actual detections. However, both estimates
are drawn from the limited sky area of this work. The original EMSS sky area included the full unobstructed
field of view of the IPC detector, whereas we have used only the inner 19 arcmin of each field. A direct
scaling of the sky areas used would be inappropriate because 
the outer regions of the detector have lower
effective exposure times, along with other effects. However, an approximation to the correct scaling is
the number of EMSS sources detected in these two different sky regions. A total of 478 of the 835 total
EMSS sources fall within the restricted survey area of this catalog. Using this scaling increases the
number of missing clusters expected in the full EMSS sky area to 
$13.1-24.9$. 

Making the assumptions that the redshift and luminosity distributions of currently undiscovered clusters
in the EMSS sky area match those we have currently discovered, we add clusters to the redshift and
luminosity bins accordingly. In Table \ref{tab_emssbins} we show the original EMSS XLF data of H92
modified to reflect the updates from more recent work described above (first entry in each bin). 
Additionally, we show the number of clusters in each bin resulting from the additions of only those
clusters we have found thus far in the catalog (Table \ref{tab_newclusters}; second entry in each bin).
The third entries, shown as a range, are our best estimate of the total number of clusters we expect to
exist in the EMSS, by adding the estimated total new cluster numbers. Note the we have added a fourth
redshift shell,
$0.60<z<0.85$, as described above; we do not combine it with the
$0.30<z<0.60$ shell in order to better compare with the H92 results. 

In order to calculate the new XLF including the addition of new clusters, we must calculate the accessible
volume ($V_a$) for each additional cluster to find its contribution to each ($z$,$\,$L$_X$) bin. Based in
part on the method detailed in H92, we calculate the luminosity function (taking care to appropriately
weight possible clusters whose estimated L$_X$ spans more than one bin). The volume $V_a$ for each cluster
must be calculated based on its detect cell flux, $F_{det}$. Because our new clusters were not detected
using the EMSS detect cell, we must estimate the value of
$F_{det}$ that the EMSS would have obtained.  Earlier (see \S \ref{subsec_sumnonrandom}) we argued that 
the 3rd ($8.4\arcmin$) aperture is  a good estimate of the total X-ray flux from these distant clusters
\citep[see][for details]{thesis}. We then assume the same conversions between $F_{det}$ and $F_{tot}$ as
used by H92 and \citet{gio90b} to calculate $V_a$ for each new cluster according to the prescription of
\citet{gio90b}, H92, and \citet{hen00}, and so obtain $N(L)$ in each bin. We note that we take into
account the IPC PSF, and the finite extent of a cluster's emission (as described in
\citealt{hen00}) in performing our corrections. This procedure slightly overestimates $V_a$ (due to a
slight overestimate of the detect cell flux for diffuse clusters; \citealt{ada00}); new X-ray images of
these clusters would be required to obtain more accurate $F_{det}$ and $F_{tot}$ values. We emphasize that
assuming canonical surface brightness profiles for the new clusters ($\beta=2/3$;
$r_{core}=250$~\hkpc), even though our detection algorithm implies they are more diffuse, conservatively
reduces their impact on the XLF described below. With accurate surface brightness profile data in hand, we
would expect to calculate lower $V_a$ values, and thus greater $N(L)$ values in each bin.

\begin{figure*}
\epsscale{2.0}
\plotone{\figdir 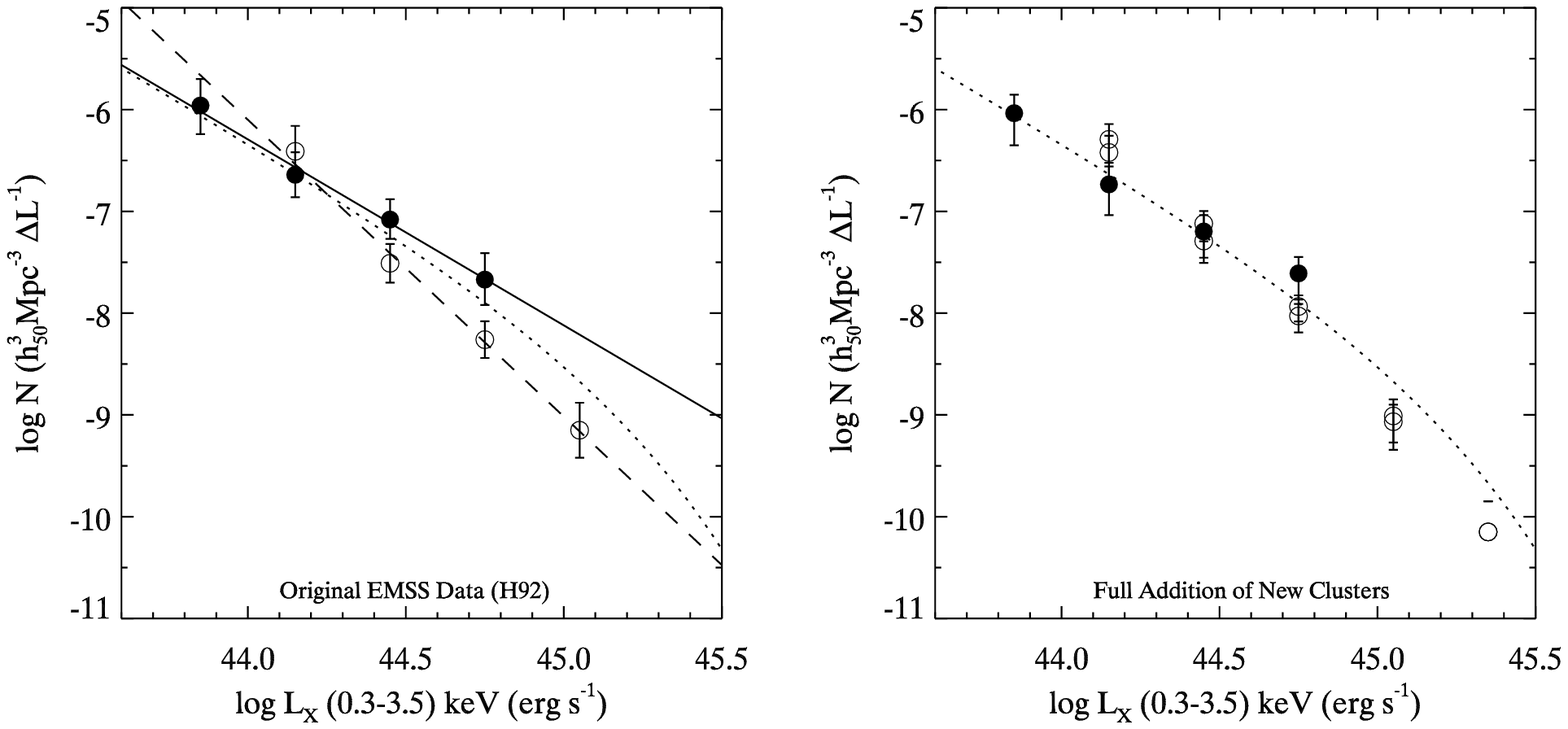}
\caption[Evolution in the cluster XLF for the Original \citet{hen92_h92} Sample vs. the Revised EMSS
Sample]{The X-ray Luminosity Function for clusters of galaxies in the EMSS sample in the first
($0.14<z<0.20$; filled circles) and third ($0.30<z<0.60$; open circles) redshift shells. The left panel
shows the original XLF determinations of H92. The right panel shows our revised values based upon the
addition of new clusters from the HFS sample after scaling for the number expected to be in the full EMSS
sky area (there are two open circles in each bin, the lower range is the sky area scaling of only those
7.5 clusters already found in the HFS sample, the upper end of the range includes the extrapolation of
clusters we would expect to find if we investigated the remainder of the HFS sample, also sky area scaled;
see Table~\ref{tab_emssbins}, and
\S~\ref{subsec_xlf}). The corrections  and updates shown  in Figure
\ref{fig_xlf_comp_update_h92} are included here (in the right panel) as well. The solid and dashed lines in
the left panel are power-law fits to the low and high-$z$ redshift shells, respectively. Errors on the data
points are the square-root of the total number of objects in each bin. The XLF from the BCS sample
\citep{ebe97} is overlaid as a dotted line.}
\label{fig_newxlfevolution}
\end{figure*}

The ``negative'' evolution of the XLF reported by H92 was based on the difference in the XLF between the
lowest and highest redshift shells (see Fig. 3, H92) primarily in the fourth luminosity bin ($44.6 \leq
$log L$_X \leq 44.9$ \lx) where the XLF data differed at the $3\sigma$ level.  In Figure
\ref{fig_newxlfevolution} we show the EMSS XLFs in the first ($0.14<z<0.20$; filled circles) and third
($0.30<z<0.60$, open circles) redshift shells. These data include all the updates to the EMSS sample since
H92 as well as the full addition of new clusters that we estimate to exist in the entire EMSS sky area
from the current work (third entries in each bin, Table
\ref{tab_emssbins}).

\begin{\myfigure}
\centerline{\epsfig{file=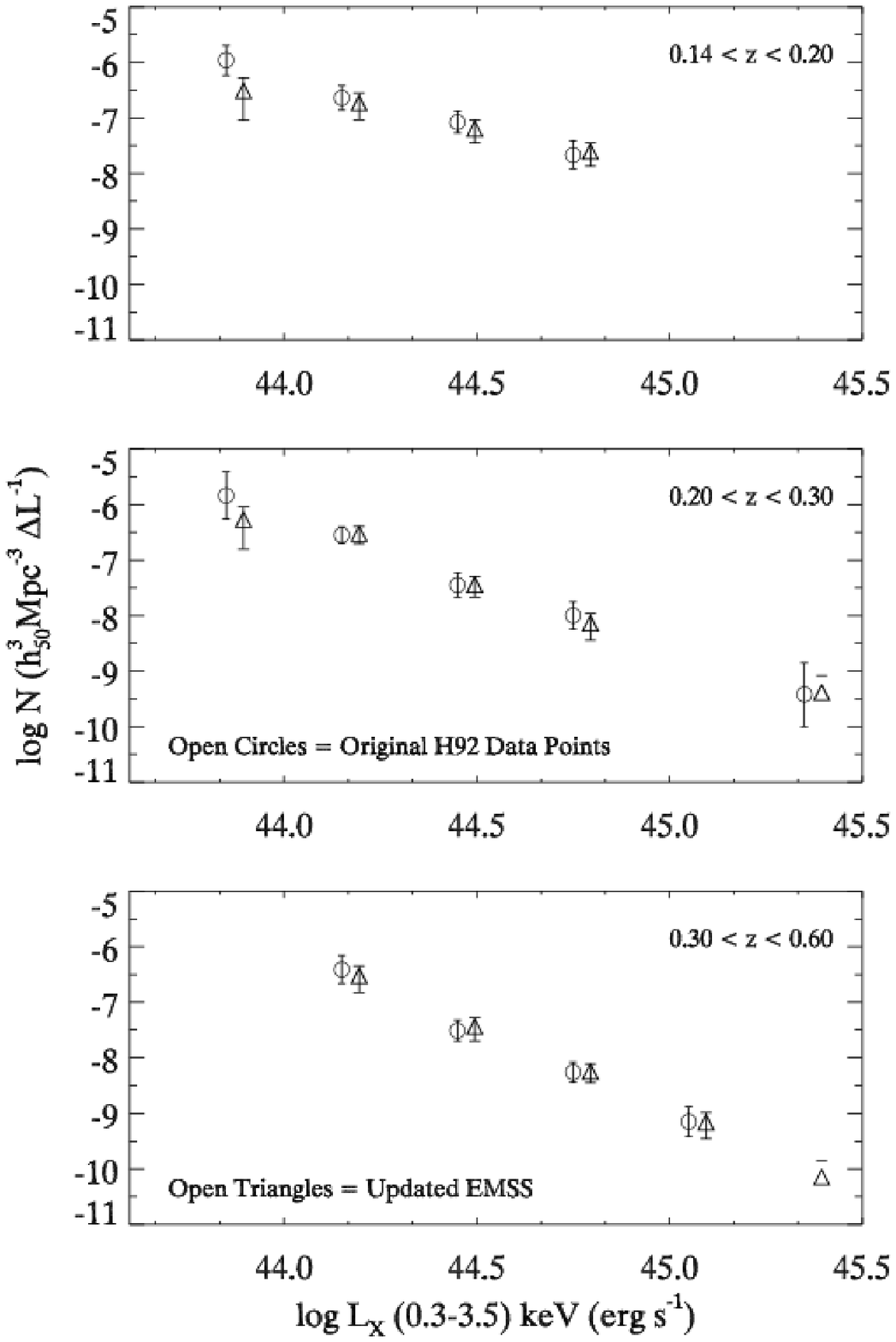, scale=0.60}}
\caption[The Updated EMSS Cluster X-ray Luminosity Function in 3 Redshift Shells]{The X-ray Luminosity
Function for clusters of galaxies in the EMSS sample in three redshift shells. Open circles are the
original EMSS values as given by H92 (taken directly from Figures 2 and 3 in that work), open triangles
indicate the values after updating for more recent work since H92 (the open triangles have been offset by
0.045 log L for clarity). Error bars are 1~$\sigma$ errors computed from the number of objects in that bin
using Poisson statistics. These values do not include the addition of any new clusters found in this work.}
\label{fig_xlf_comp_update_h92}
\end{\myfigure}
\smallskip
\epsscale{1.0}

The addition of only a few clusters at high redshift and luminosity, both due to new data on existing EMSS
clusters and to the estimated number of new clusters from this work, just barely allows the error bars in
all the mutual L$_X$ bins to overlap at the $1\sigma$ level. So we have reduced the strongest evidence of
evolution (in the fourth L$_X$ bin) to only $1\sigma$. In order to make a better estimate of evolution in
the XLF, we have overlaid the XLF derived from the
$z\leq0.3$ Bright Cluster Sample (BCS) of
\citet{ebe97} as a dotted line in both panels of Figure \ref{fig_newxlfevolution}. We can see that in the
right panel (new data) the fourth luminosity bin ($44.6 \leq $log L$_X \leq 44.9$
\lx) is now entirely consistent with the BCS XLF. The fifth and sixth bins do show a small deficit of
clusters relative to the BCS. Taking into account the error bars from the BCS, the two samples differ only
at the
$1\sigma$ level in any individual bin. Thus, we do not require evolution in the XLF at any X-ray
luminosity. The deficit of clusters between our estimated high-$z$ EMSS XLF and the BCS XLF is only 5 and
2 clusters in the fifth and sixth L$_X$ bins, respectively. Given that our estimates of new clusters in
the EMSS are conservative in basing total additions solely on the HFS subsample, seven additional clusters
in the full catalog would not be surprising. These results differ from the recently renewed measurements
of high-L$_X$ evolution found by some other groups \citep[see][for a thorough summary]{gio01}.

\epsscale{1.0}

\section{Discussion and Conclusions\label{sec_concl}}

In summary, a re-analysis of the complete {\it Einstein} IPC image data with improved detector-response
information and a multi-aperture detection algorithm has generated a catalog of 6610 X-ray sources. Of
these, 772 sources are detected within a subset of the EMSS sky area. The vast majority of these
potential new sources (97\% of the random subsample)  would have had insufficient flux in an EMSS-like
detection aperture to merit inclusion in the original catalog and were detected here by using larger
apertures.  A statistical analysis of the relative fluxes in the four apertures used in the source
detection algorithm indicates that these new sources appear more spatially extended than the original EMSS
sources. There are several possible explanations for why these sources were not included in the original
EMSS: (1) improved flat fielding has reduced systematic errors in the detection process, resulting in the
recovery of additional EMSS-like sources; (2) these sources tend to have softer X-ray spectra (e.g., those
of stars) which emit primarily at lower energies where the width of the X-ray telescope-IPC point response
function is larger and makes these sources appear more extended and thus less likely to have been detected
in the EMSS;  (3) some of these sources are related to targets, are affected by the telescope ribs, or
have S/N values that are increased by contributions from non-EMSS exposures; (4) many of these sources are
actually physical or projected aggregates of sources which individually are below the flux limits of the
EMSS but have the necessary ensemble flux at angular scales corresponding to the large apertures used
here; and (5) these sources are actually spatially extended objects.   Only explanation (2) can be
eliminated from being the reason for many of these new sources. The spectral characteristics of the {\it
ROSAT} PSPC counterparts to some of these sources do not support the hypothesis that the majority of these
sources are spectrally soft, nor have we identified many of these sources as being due to stars. All of
the other explanations appear to contribute substantially to the catalog.

Primarily to distinguish between explanations (4) and (5), we scrutinized a random subsample of new
sources. In a randomly-selected subset of 133 IPC fields there are 73 new serendipitous source candidates
(compared with 49 original EMSS sources within the same area of sky), which have been investigated through
database searches and optical imaging, revealing that the vast majority of sources are not clusters of
galaxies.  One possible new distant cluster ($z=0.52-0.59$) and one nearby group ($z=0.018$) were found.
The possible distant cluster detection in this sample implies that $\leq2.3\%$ (0--1 of 43 random sources)
of the catalog are distant clusters.

Specifically targeting different subsets of the catalog, which we pre-selected to be more likely clusters
of galaxies, we searched available databases, the literature, and {\it ROSAT} surveys from other groups
for cluster identifications. In addition, we conducted an imaging survey of 34 pre-selected sources. This
investigation yielded 11 distant ($z>0.14$) clusters (3 of which are only possible cluster
identifications). Over half of these clusters are confirmed as X-ray sources by available {\it ROSAT} PSPC
and HRI images. These clusters are missing from the EMSS, and many of them have large estimated galaxy
overdensities consistent with high X-ray luminosities. Having confirmed the existence of a modest number
of rich, distant clusters in the catalog, we further scrutinized the catalog to create a comparison sample
as accurately matched as possible to the EMSS selection criteria.  This allowed us to estimate
statistically the EMSS catalog incompleteness. Using only the  newly discovered clusters as a minimum, and
our discoveries scaled to the full EMSS sky area catalog as a maximum, we estimate the EMSS cluster
catalog of H92 to be $72-83\%$ complete. These new clusters are plausibly lower in X-ray surface
brightness than previous EMSS detections and could be clusters still in the process of virializing. As
such, the examples found here may contain somewhat different galaxy populations than those found in X-ray
luminous clusters heretofore
\citep[e.g.,][]{ell01}. Source \#420, for example, is clearly asymmetric and spatially extended, though it
lies at very high redshift, according to the detailed analysis of its {\it ROSAT} PSPC image by \citep[see
also Appendix \ref{sec_appendix}]{ebe00}. Furthermore, the measured X-ray temperature of source \#420 is
$6.46^{+1.74}_{-1.19}$~keV \citep{del00}, approximately half the value found for MS~$1054.4-0321$
\citep{don98}, a cluster with nearly identical redshift and X-ray luminosity, suggesting that these two
clusters are in very different dynamical states. Therefore, as shown in \citeauthor{del00}, the L$_X-$T$_X$
relationship must have considerable spread at $z \sim 0.8$, as would be expected if the evolutionary
status of clusters at that epoch is much more diverse than today. In addition, the specific sources
\#1492, \#1605, and \#2436 are our best examples of such clusters since these clusters appear
significantly richer in galaxies than their X-ray luminosity would indicate (see Table
\ref{tab_newclusters}). Additional details of the optical imaging program and the optical properties of
these new clusters can be found in Paper 2.

We then estimated the effect that these missing clusters would have on the EMSS XLF and its evolution.
First, we have made corrections to the original EMSS cluster identifications, redshifts, and other
particulars based upon new information available since H92.  Then we add the newly discovered clusters to
the sample, recalculating the XLF in the same redshift shells used by H92. While there are changes to
nearly all the redshift and luminosity bins, the majority of cluster additions occur in the intermediate
and high$-z$, high$-$L$_X$ bins. Our additions have reduced the deficit of high redshift clusters in the
fourth luminosity bin to $1\sigma$, nearly removing the strongest evidence for evolution in the EMSS
cluster XLF. Comparing to the BCS cluster sample, we do not require evolution in the XLF within any
luminosity bin for consistency at high and low redshift at the $1\sigma$ level. Our expectation for the
number density of clusters in the EMSS at
$0.3\leq z \leq 0.6$ is lower than the BCS values at $z\leq0.3$ in the fifth and sixth luminosity bins,
equivalent to a deficit of only 5 and 2 clusters respectively. Because our estimates for the number of
clusters missing from the EMSS were conservative, it is plausible  the high-$z$ EMSS XLF is entirely
consistent with the BCS XLF. We caution, however, that the full number of new clusters in this sample is
not known to great precision, and further optical imaging of these sources is required to be certain. The
cosmological constraints from our result are complex, and without accurate X-ray temperatures, or other
mass estimators for the new clusters, we cannot be definitive.  However, most cosmological models
constrain luminosity evolution in clusters and the basic trend is that for small values of
$\Omega_{\rm matter}$, clusters evolve slowly, and for high values they evolve more rapidly \citep[see
e.g.,][]{edg90}.  Our addition of clusters mildly decreases the existing evolutionary constraint, which
when combined with mass estimators, sets an upper limit to
$\Omega_{\rm matter}$, slightly decreasing the allowed values from the limits set by
\citet{don99a} of
$\Omega_{\rm matter} <0.45\pm0.1$ for an open Universe, and $\Omega_{\rm matter} <0.29\pm0.1$ for a flat
Universe. Future X-ray temperature measurements with XMM$-${\it Newton} as well as optical velocity
distribution data for these clusters will solidify these new constraints.

While the EMSS cluster sample suffers significantly from the use of a single, fixed detect cell size, it
also has substantial advantages over other, more recent samples, which warrant its continued
investigation: (1) the EMSS source selection is not based upon the X-ray source being resolved as with
newer surveys \citep[e.g., WARPS, SHARC,][]{ros98} such that highly concentrated X-ray clusters (if they
exist) would not be misidentified \citep[see][]{don01}; (2) virtually all EMSS sources have been optically
identified (we do note that the large sky-area NEP sample is also virtually completely identified;
\citealt{gio01}), and indeed, even further scrutinized
\citep[e.g., the work of][has found several sources originally identified as low (L$_X$,$z$) clusters that
are actually BL Lac objects]{rec99} making the optical identifications more secure; and (3) many
IPC-detected clusters were reobserved, either as pointed targets or ``serendipitously'', allowing a more
detailed surface brightness analysis than is possible for new {\it ROSAT} detections. This latter
advantage allows a much more secure correction from observed flux to total flux than for other samples and
will be used to provide a final EMSS XLF determination in the third paper in this series
\citep{P4}.

Finally, we emphasize that, despite the uncertainties in this analysis due to the incomplete optical
identification of the full 772 source catalog, enough bona fide new distant clusters were discovered to
cast doubt on the evidence for evolution in the cluster XLF out to
$z\sim0.5$. Further, the reason for this new conclusion is the recognition that the original EMSS cluster
sample is surface brightness, not flux-limited, and missed several high-$z$, high-L$_X$ clusters due to
this selection bias. And while the EMSS detection methodology may be the most susceptible to this bias,
other cluster detection techniques (e.g., wavelet, VTP) probably have this bias present to a lesser degree
\citep[see][]{ada00}. Therefore, it cannot be assumed that {\it any} current X-ray discovery technique has
detected all low surface brightness clusters above their stated flux limit.

\myacknowledgments

The authors wish to thank all those who shared data and information with us, including Tadayuki Kodama,
Isabella Gioia, Eric Perlman, and Kathy Romer, and especially an anonymous referee who pointed out a
significant error in our assumed maximum redshifts, leading to an important correction to our results. ADL
wishes to thank the KPNO support staff during the observing runs for this project, Michael Harvanek for
help with data analysis and reduction, the HEASARC facilities for maintaining invaluable research tools and
databases, and Beth White for her ongoing support. ADL made extensive use of the relations summarized in
\citet{hog99}, and gratefully acknowledges that author. ADL and this work were supported by a NASA
Astrophysical Data Program grant
\#NAG5-6936 and by travel grants for thesis work at Kitt Peak National Observatory by NOAO. EE
acknowledges support provided by the National Science Foundation grant AST 9617145. EJG wishes to thank
David Helfand for motivating and guiding the original construction of the IPC source catalog and Ben
Oppenheimer for generating the original IPC source catalog used as the basis for this work. EJG
acknowledges support from NASA contract NAS8-38249 and grant NAC5-1656 at MIT.  We gratefully acknowledge
Dr. Harald Ebeling for a detailed criticism of the original manuscript which led to a significant revision
of this research.  This research has made use of the High Energy Astrophysics Science Archive Research
Center (HEASARC) provided by the NASA-Goddard Space Flight Center, the Einstein Online (EINLINE) database 
and the NASA Astrophysics Data System (ADS) maintained at the Smithsonian Astrophysical Observatory,  the
NASA-IPAC Extragalactic Database (NED) operated by the Caltech Jet Propulsion Laboratory under contract
with the National Aeronautics and Space Administration, and the SIMBAD database, operated at the Centre
des Donnees Astronomiques, Strasbourg, France.  The Digitized Sky Surveys were produced at the Space
Telescope Science Institute under U.S. Government grant NAG W-2166. The images of these surveys are based
on photographic data obtained using the Oschin Schmidt Telescope on Palomar Mountain and the UK Schmidt
Telescope.



\begin{appendix}

\section{Appendix: Notes on Individual Sources \label{sec_appendix}}

In this appendix we provide additional detail for individual sources of interest. We describe objects
excluded from our catalog under special circumstances, sources in the three subsamples which were judged
not to be groups or clusters of galaxies, and clusters discovered by other research groups. Those clusters
discovered in the imaging program of this work will be presented in full detail in Paper 2. To obtain
unabsorbed flux from instrumental count rates (and convert between different bandpasses used by different
groups) we have assumed a power-law spectral energy distribution with photon index $\Gamma = 1.5$ ($\alpha
= 0.5$), and used the WPIMMS software to calculate conversions, using the weighted average neutral
hydrogen column density obtained from W3nH. We assume the same spectral
energy distribution to perform K-corrections when calculating rest-frame X-ray luminosities. These notes
are appended to illustrate the scrutiny to which potential cluster sources were subjected by this work.

\subsection{Sources \#324 \& \#962}
Detected in the HFS subsample, these two clusters were originally found by the first installment of the
EMSS \citep[the Medium Survey \#1, MSS1 hereafter,][]{mac82,sto83} but were not members in the final EMSS
Catalog, as noted in
\citet{gio90a}, because some of the integration time was discarded due to a more conservative detection
algorithm than was used for the MSS1. Thus, these two source fell below the 4$\sigma$ detection limit of
the EMSS. Why these sources are redetected here is not completely obvious but, since both are identified
as clusters, we assume that it is due to their extended flux and have retained them in the catalog. Source
\#324  (R.A., Decl. (J2000) $= 01^h29^m02\fs06, +~07\arcdeg40\arcmin54\farcs7$)
is identified with a compact galaxy group known as Shakbazian 41 and is of individual interest due to the
significant signs of interaction among its brightest member galaxies. Source \#962 lies at R.A., Decl.
(J2000) $= 04^h41^m41\fs53, -~10\arcdeg56\arcmin47\farcs0$.
A third new
catalog source (\#4402, R.A., Decl. (J2000) $= 09^h41^m04\fs84, +~11\arcdeg36\arcmin31\farcs1$), while not
a member of the HFS subsample, was also detected in MSS1, where it was identified as a star. However, this
source is also in the vicinity of a small, nearby galaxy group cataloged recently by
\citet{ram97} within the CfA redshift survey sky area. In this paper we identify this source with this
group because the current technique found it to be extended, although only new, sensitive X-ray
observations can determine the true identity of source
\#4402. All three MSS1 sources have quite modest X-ray luminosities, consistent with being small groups or
poor clusters of galaxies. 

\subsection{Source \#420}
Re-detected in the HFS subsample of our catalog, this distant cluster ($z=0.833$, R.A., Decl. (J2000) $=
01 ^h 52 ^m 43 \fs 35$, $-~13 \arcdeg 58 \arcmin 01 \farcs 5$) has been rediscovered in the {\it ROSAT}
PSPC database independently by three different groups
\citep{ros98,ebe00,rom00} and its reason for not being discovered in the
EMSS has been discussed in detail by
\citet{ebe00}. Briefly, \citeauthor{ebe00}{} argue that the asymmetric nature of the X-ray emission
from this cluster caused the EMSS to mis-locate an accurate centroid and so underestimate the flux
due to its asymmetric and somewhat low surface brightness nature. 
We have not obtained optical images for this cluster but the extent
seen by {\it ROSAT} as well as the temperature and Fe line measured by {\it Beppo} SAX are ample
reasons that the cluster ID is secure. It has L$_X = 8\times10^{44}$
\lxh{} in the $0.5-2.0$~keV band \citep{ebe00}, which corresponds to $14.6\times10^{44}$
\lxh in the \eband{} band,
and an unabsorbed flux of
$5.00\times 10^{-13}$ \fx. The first aperture IPC
unabsorbed flux for this source is
$3.61 \times 10^{-13}$ \fx, but it increases to $6.9$ and $10.5
\times 10^{-13}$ \fx{} in the second and third apertures, respectively. The X-ray
contour maps of this cluster shown in \citeauthor{ebe00} indicate a spatial extent of at least 5
arcmin in diameter, suggesting that at least the flux in the 2nd aperture should be used. We
therefore regard the luminosity value of \citeauthor{ebe00} as a lower limit, but we will adopt it
instead of our own estimates because \citeauthor{ebe00} have performed a more detailed analysis of
the newer available X-ray data.

\subsection{Source \#992}
This field, part of the ``ramp'' subsample, has several identified galaxies from the nearby cluster
Abell 516 at
$z=0.1407$, which has an optical center 7\farcm1 away from the X-ray source location, based on the
SIMBAD coordinates. However, upon a more detailed investigation, we found that the study which
obtained the optical galaxy redshifts \citep{cia85} notes that the optical cluster center appears to
be in the vicinity of the two galaxies with measured redshifts and thus at the location of source
\#992. A visual inspection of the SIMBAD position for Abell 516 on the DSS finds no large
concentration of galaxies and that the nearest concentration of galaxies is at the source \#992
position. Therefore, we assign source \#992 to Abell 516. Further, the target of the original IPC
image in which this source was discovered is a supercluster which includes Abell 516. Therefore,
source \#992 is eliminated from the catalog as target related for the same reason as source \#4359
in the random source subsample. Without the extensive analysis provided by the inclusion of this
source in the non-random ``ramp'' sample, this source would have remained in the catalog as a viable
serendipitous source (see \S \ref{subsec_moreacc}). 

\subsection{Source \#1757}
A member of the random sample, this source exhibits a small concentration of galaxies in its
color-magnitude diagram consistent with a group at $z\sim0.35$, but the formal measured B$_{gc}$ value is
negative. A {\it ROSAT} All-Sky-Survey source lies 4.2 arcmin NNW of \#1757 and so could only
contribute flux to the third aperture. We conclude that this cluster is not rich enough and so
would not be bright enough to have this source be identified as a cluster of galaxies. It is more
likely a blend of sources.

\subsection{Source \#1767}
A member of the HFS subsample, we initially identified this source as a cluster based on a large galaxy
over-density, a good concentration in its color-magnitude diagram at $z\approx0.4$, and a measured
B$_{gc}$ value of $\approx 1000$~\bggmph. The X-ray centroid lies at 
R.A., Decl. (J2000) $= 10 ^h 48 ^m 28 \fs 84, +~06 \arcdeg 43 \arcmin 26 \farcs 8$.
However, we found a RASS FSC detection lying within the 2nd IPC
aperture, but $>500$~\hkpc{} away from the BCG of the apparent cluster. This source was
positionally consistent with a prominent galaxy which is nearly a full magnitude brighter in $B$ than two
adjacent galaxies, though all three had similar $R$ and $V$ colors. The galaxy also appeared more
centrally concentrated in $B$, suggesting it is an AGN. Combined with the fact that the FSC source was
of equivalent X-ray flux to our measurement of the second IPC aperture flux, it seems very likely that
this AGN contributes significantly to the X-ray detection. Even if the apparent cluster is
as rich as measured, at best this field is a blend of sources, neither of which would be of sufficient
individual flux to be added to the EMSS. This source only obtains a S/N $>4.0$ in the third and fourth IPC
apertures, further suggesting that is a blend.

\subsection{Source \#1772}
Also a member of the HFS subsample, this field contains an apparent overdensity
of galaxies. The X-ray centroid lies at 
R.A., Decl. (J2000) $= 10 ^h 49 ^m 48 \fs 98, +~06 \arcdeg 44 \arcmin 53 \farcs 6$.
However, we find that a large fraction are late-type galaxies, and that there are no clumps in
color to indicate a significant galaxy overdensity (i.e., we find B$_{gc} \leq 250$~\bggmp,
which is below Abell Richness Class 0) in any part of the image . Unfortunately, the exposures in one
of our filters are not photometric for this field, and galaxy color measurements may be inaccurate.
However, it seems to be an optical superposition of many field galaxies, with some possible poor groups as
well. Therefore we do not identify this field as a cluster of galaxies.

\subsection{Source \#2036}
Also a random sample source, this field has two apparently overlapping concentrations of galaxies.
The first group has an estimated galaxy over-density of B$_{gc}\sim50-200$~\bggmph, at redshift
$z\sim0.35-0.45$ which suggests an X-ray luminosity of L$_X = 0.004-0.1\times10^{44}$ \lx{} and
an expected X-ray flux of $0.01-0.15 \times 10^{-13}$ \fx{} \ebandp. The
second group has an estimated galaxy over-density of B$_{gc}\sim300$~\bggmph, at redshift $z\sim0.60$
which suggests an X-ray luminosity of L$_X = 0.3\times10^{44}$ \lx{} and an expected
X-ray flux of $0.12 \times 10^{-13}$ \fx{} \ebandp. The observed first aperture
flux of this source is $1.0 \times 10^{-13}$ \fx{} \ebandp, significantly higher
than even the combination of both possible groups. Additionally, no PSPC source was detected in the region
of the X-ray centroid or the galaxy overdensity to a 3$\sigma$ limit of $1 \times 10^{-13}$ \fx{} \ebandp,
although a 6$\sigma$ PSPC source lies 4 arcmin to the south of \#2036 (see  Table
\ref{tab_xraydbrandom}) and could contribute flux to the third aperture detection of this source. We
conclude that, while the ID of this source is not obvious, it is not a cluster of galaxies, but a
blend of sources, similar to source \#1757.   

\subsection{Source \#2203}
This is a $z=0.39$ cluster identified in the SHARC \citep{rom00} as
RXJ1241.5+3250. Our X-ray centroid lies at 
R.A., Decl. (J2000) $= 12 ^h 41 ^m 33 \fs 56, +~32 \arcdeg 49 \arcmin 50 \farcs 4$.
Based on the provided
{\it ROSAT} PSPC count rate of 0.04
cts s$^{-1}$ in the $0.4-2.0$~keV band \citep{rom00} we calculate an
unabsorbed flux of $4.66 \times 10^{-13}$ \fx~($0.4-2.0$~keV), equivalent to $8.71
\times 10^{-13}$ \fx{} in the \eband{} band. The unabsorbed fluxes of
$8.52$ and $11.8 \times 10^{-13}$ \fx{} in the 2nd and 3rd IPC apertures, respectively,
are consistent with the SHARC estimates, and confirm their cluster ID.
We adopt the redshift and X-ray
luminosity values from their work.

\subsection{Source \#2407}
Another HFS subsample member, this field contains an apparent optical overdensity of galaxies.
The X-ray centroid lies at 
R.A., Decl. (J2000) $= 13 ^h 19 ^m 36 \fs 59, +~55 \arcdeg 06 \arcmin 35 \farcs 6$.
However, under detailed scrutiny we find only two small physical concentrations. The first clump to the NNE
is at a redshift of $z\sim0.20-0.25$, with galaxy overdensity of B$_{gc} \approx210$~\bggmph, and an
approximate luminosity of only L$_X = 9\times10^{42}$ \lx, corresponding to f$_X = 1.8\times10^{-14}$ \fx.
The second clump to the South is at a redshift of
$z\sim0.35-0.40$, with galaxy overdensity of B$_{gc} \approx 250$~\bggmph, and thus L$_X =
1.5\times10^{43}$ \lx, corresponding to f$_X = 1.3\times10^{-14}$ \fx. The smallest
IPC aperture detection corresponds to $4\times10^{-13}$ ergs cm$^{-2}$ s$^{-1}$, easily greater than the
sum of these sources. We must assume that an as yet unidentified AGN in the field is the bulk of our IPC
detection. Therefore, we do not identify this source as a cluster of galaxies.

\subsection{Source \#2616}
This source is identified as an extended X-ray source by \citet{vik98a}. However, they
cannot make an optical confirmation of a cluster due to a nearby bright star obscuring the field.
Our X-ray centroid lies at
R.A., Decl. (J2000) $= 14 ^h 15 ^m 40 \fs 29, +  19 \arcdeg 06 \arcmin 00 \farcs 5 $.
We leave this field unidentified.

\subsection{Source \#2626}
This is a $z=0.138$ cluster identified in the SHARC \citep{rom00} as
RXJ1416.4+2315. 
Our X-ray centroid lies at
R.A., Decl. (J2000) $= 14 ^h 16 ^m 27 \fs 35, +  23 \arcdeg 15 \arcmin 24 \farcs 6 $.
Based on the provided
{\it ROSAT} PSPC count rate of 0.112 cts s$^{-1}$ in the $0.4-2.0$~keV band \citep{rom00} we calculate an
unabsorbed flux of $13.3 \times 10^{-13}$ \fx{} ($0.4-2.0$~keV), equivalent to $24.8
\times 10^{-13}$ \fx{} in the \eband{} band. The unabsorbed fluxes of
$10.9, 19.9,$ \& $38.8 \times 10^{-13}$ \fx{} in the 2nd, 3rd, \& 4th IPC apertures,
respectively, are consistent with the SHARC estimates, and confirm their cluster ID.
We adopt the redshift and X-ray
luminosity values from their work.

\subsection{Source \#2937}
This field may contain a loosely concentrated poor group at low redshift ($z<0.10$), judging by the
appearance of nearby galaxies in the field. Our X-ray centroid lies at R.A., Decl. (J2000) $= 15 ^h 36 ^m
02 \fs 20, +  23 \arcdeg 18 \arcmin 34 \farcs 3 $. The low X-ray luminosity of such a group makes it
unlikely to be the dominant X-ray emitter in this field, if it is a physical association of galaxies at
all. We can only state that this is not a rich cluster of galaxies, and that we do not as yet identify an
obvious optical counterpart to the X-ray emission.

\subsection{Source \#3065}
This field contains an apparent distant cluster $6\arcmin$ SW of the field
center in a deep $R$ image. Our X-ray centroid lies at
R.A., Decl. (J2000) $= 16 ^h 08 ^m 53 \fs 06, +  28 \arcdeg 53 \arcmin 31 \farcs 3 $.
Unfortunately, we do not have the multi-color data necessary to estimate its redshift and richness. 
However, because the cluster is so far from the X-ray centroid, we consider it likely that this field
contains a blend of sources, and the cluster only contributes to the 3rd and 4th aperture detections, which
exhibit a dramatic increase in S/N compared to the second aperture. Therefore we do not identify this
source as a cluster of galaxies, but leave it unidentified pending more observations.

\subsection{Source \#3175}
This field is identified as J164154.2+4000033, a $z=1.005$ QSO found by
\citet{cram92}. \citet{vik98a} identify this source as a cluster of galaxies at
$z=0.44-0.55$. It seems likely that this field appears as a blend of sources to the IPC, and so we do not
add it to our cluster list.

\subsection{Source \#3469}
This source is identified by \citet{vik98a} as a portion of Abell S840 ($z=0.0152$). We concur and add it
to our cluster identifications.  Our X-ray centroid lies at R.A., Decl. (J2000) $= 20 ^h 03 ^m 28 \fs 38,
-  55 \arcdeg 56 \arcmin 44 \farcs 8 $. Based on their provided flux of $4.76\times10^{-13}$
\fx{} ($0.5-2.0$~keV), we adopt a luminosity of L$_X = 41.95\times10^{44}$ \lxh{} in the \eband{} band.

\subsection{Source \#4057}
This field, which is part of the random sample, exhibited no obvious galaxy overdensity and so was not
evaluated for cluster richness. Our X-ray centroid lies at R.A., Decl. (J2000) $= 02 ^h 39 ^m 51 \fs 80,
-  23 \arcdeg 20 \arcmin 42 \farcs 8 $.  Our optical imaging indicates only a very weak galaxy overdensity
at best. However, this source is identified as a
$z=0.42-0.53$ cluster by
\citet{vik98a}, although they do not have spectroscopic confirmation. The {\it ROSAT} extrapolated total
flux (including absorption by galactic hydrogen) for the purported cluster is given by
\citeauthor{vik98a} as $8.4 \pm 1.8 \times 10^{-14}$ \fx{} in the $0.5-2.0$~keV bandpass. From the IPC
data, we estimate the absorbed flux in the same bandpass to be $1.1$ and $2.4\times 10^{-13}$
\fx{} in the first and third apertures, respectively. The S/N rises from 3.5 to 4.4 between these same
apertures, so source \#4057 enters the catalog due to its third aperture flux. However, the PSPC
observation also detected a point source with flux $1-2\times 10^{-13}$ \fx{} 3.3 arcmin from the IPC
source location, and thus contributing substantially to the third aperture flux. Therefore, we identify
source
\#4057 as a combination of sources and so do not identify the possible cluster of galaxies of
\citeauthor{vik98a} as responsible for the entry in our catalog.

\subsection{Source \#4746}
This field is apparently an optical superposition of galaxies with no discernible concentration in
redshift.  Our X-ray centroid lies at R.A., Decl. (J2000) $= 13 ^h 34 ^m 03 \fs 27, -  08 \arcdeg 26
\arcmin 14
\farcs 4 $.  The most significant structure in the color-magnitude diagram lies at approximately
$z\sim0.13-0.17$, but our range of estimates for the B$_{gc}$ value includes zero. This may be a very poor
group, but it is not responsible for the X-ray emission detected in this field, which is likely to be a
blend of sources. We note that \citet{vik98a} identify this field as an extended X-ray source, and a
``likely false'' cluster of galaxies, because it did not appear as a cluster in their optical imaging. We
also do not identify this source as a cluster of galaxies.

\end{appendix}

\begin{\mydeluxetable}{lrrccccrrrc}
\tablecaption{New Sources in the Randomly-Selected Subset of 133 EMSS IPC Fields
\label{tab_randomsample}}
\tablewidth{0pt}
\tablehead{
\colhead{Cat. \#} & \colhead{RA J2000} &
\colhead{Dec. J2000} & \colhead{Seq.} &
\colhead{$c.r._1$\tablenotemark{a}} &
\colhead{$c.r._2$\tablenotemark{a}} & \colhead{$c.r._3$\tablenotemark{a}} &
\colhead{$\sigma_1$\tablenotemark{b}} & \colhead{$\sigma_2$\tablenotemark{b}} &
\colhead{$\sigma_3$\tablenotemark{b}} & \colhead{Eval.\tablenotemark{c}}
}
\startdata
64	&	00	25	52.52	&	+	17	17	30.2	&	1810	&	4	&	10	&	16	&	3.74	&	4.52	&	5.09	&	X	\\
69	&	00	26	47.83	&	+	17	22	59.9	&	1810	&	3	&	5	&	19	&	3.02	&	2.78	&	4.47	&	X	\\
71	&	00	27	35.88	&	+	17	06	25.7	&	1810	&	5	&	11	&	31	&	3.53	&	3.71	&	6.29	&	X	\\
97	&	00	38	37.56	&	+	29	32	05.3	&	7917	&	3	&	7	&	21	&	2.92	&	2.71	&	4.39	&	C	\\
3886	&	00	41	53.22	&	$-$	01	39	40.0	&	5393	&	6	&	11	&	15	&	4.42	&	4.80	&	4.90	&	X	\\
3909	&	00	48	05.59	&	$-$	25	24	43.0	&	2082	&	5	&	7	&	14	&	3.82	&	3.84	&	4.27	&	X	\\
3915	&	00	52	33.50	&	+	01	48	37.3	&	8455	&	4	&	8	&	13	&	3.56	&	4.10	&	4.04	&	X	\\
161	&	00	53	10.38	&	+	01	52	25.3	&	8455	&	4	&	8	&	15	&	3.18	&	3.23	&	4.70	&	C:	\\
163	&	00	53	18.24	&	+	01	39	20.6	&	8455	&	5	&	6	&	16	&	4.11	&	3.73	&	5.17	&	X	\\
5341	&	01	10	29.63	&	+	39	17	02.0	&	8464	&	5	&	7	&	0	&	4.50	&	4.14	&	0	&	X	\\
288	&	01	21	33.29	&	$-$	03	28	43.4	&	7208	&	4	&	8	&	13	&	3.17	&	3.55	&	4.53	&	\nodata	\\
298	&	01	23	04.75	&	$-$	03	21	59.7	&	7208	&	4	&	7	&	14	&	3.18	&	3.51	&	4.80	&	X	\\
302	&	01	23	33.74	&	$-$	03	46	29.7	&	7208	&	4	&	14	&	29	&	2.71	&	4.28	&	5.72	&	X	\\
299	&	01	23	37.36	&	$-$	03	36	55.0	&	7208	&	4	&	6	&	12	&	3.35	&	2.99	&	4.24	&	X	\\
353	&	01	37	19.54	&	$-$	04	50	08.5	&	863	&	12	&	17	&	41	&	3.05	&	3.22	&	4.58	&	X	\\
451	&	02	06	33.49	&	$-$	37	37	33.4	&	5388	&	5	&	11	&	23	&	3.29	&	3.25	&	4.55	&	\nodata	\\
446	&	02	06	37.13	&	+	23	30	48.5	&	852	&	5	&	7	&	14	&	3.74	&	4.40	&	5.29	&	X	\\
450	&	02	06	56.27	&	$-$	37	59	07.6	&	5388	&	9	&	14	&	23	&	3.80	&	4.15	&	4.86	&	\nodata	\\
4040	&	02	24	32.49	&	+	07	06	45.9	&	3256	&	0	&	6	&	23	&	0	&	3.23	&	4.96	&	X	\\
4057	&	02	39	51.80	&	$-$	23	20	42.8	&	2014	&	7	&	0	&	16	&	3.51	&	0	&	4.41	&	X	\\
657	&	03	20	05.65	&	$-$	43	10	47.7	&	3105	&	8	&	9	&	29	&	3.26	&	3.03	&	4.08	&	X	\\
804	&	03	58	55.38	&	+	10	19	44.3	&	6311	&	4	&	7	&	16	&	3.12	&	3.53	&	4.97	&	X	\\
807	&	04	00	16.33	&	$-$	36	33	16.3	&	4577	&	0	&	13	&	31	&	0	&	3.07	&	5.07	&	\nodata	\\
1415	&	08	26	10.41	&	+	26	43	12.9	&	5929	&	0	&	0	&	12	&	0	&	0	&	4.27	&	X	\\
1417	&	08	26	14.37	&	+	10	45	44.3	&	5125	&	0	&	9	&	25	&	0	&	2.76	&	4.70	&	X	\\
1424	&	08	27	23.02	&	+	26	37	15.3	&	5929	&	3	&	7	&	12	&	3.83	&	4.36	&	5.14	&	X	\\
4359\tablenotemark{d}	&	08	28	41.60	&	+	65	43	12.1	&	305	&	4	&	6	&	19	&	3.34	&	2.82	&	4.94	& 	C\tablenotemark{d}	\\
4374	&	08	47	06.81	&	+	18	34	10.4	&	4059	&	0	&	0	&	31	&	0	&	0	&	4.60	&	X	\\
1477	&	08	47	10.46	&	+	18	20	22.8	&	4059	&	7	&	0	&	23	&	3.04	&	0	&	4.02	&	X	\\
4375	&	08	52	16.54	&	+	28	27	33.8	&	5504	&	2	&	4	&	10	&	2.97	&	3.61	&	4.54	&	X	\\
1568	&	09	23	24.81	&	+	34	20	31.1	&	2101	&	4	&	5	&	18	&	3.11	&	2.93	&	4.46	&	X	\\
1584	&	09	29	31.23	&	+	06	24	27.9	&	10382	&	4	&	6	&	18	&	4.42	&	3.63	&	5.72	&	X	\\
1587	&	09	29	33.22	&	+	06	04	25.0	&	10382	&	2	&	4	&	12	&	2.74	&	2.84	&	4.41	&	X	\\
1757	&	10	44	26.87	&	+	06	32	15.6	&	6344	&	8	&	16	&	21	&	3.27	&	3.89	&	4.75	&	X	\\
4508	&	11	41	40.31	&	+	34	08	57.0	&	3530	&	0	&	10	&	26	&	0	&	2.62	&	4.65	&	\nodata	\\
1918	&	11	47	16.73	&	+	00	33	32.5	&	7712	&	9	&	15	&	22	&	3.33	&	3.94	&	4.80	&	\nodata	\\
4524	&	11	48	23.85	&	+	00	43	01.4	&	7712	&	5	&	9	&	17	&	3.05	&	3.33	&	4.63	&	\nodata	\\
1955	&	11	58	55.30	&	+	32	23	00.4	&	443	&	20	&	34	&	42	&	3.34	&	4.23	&	4.78	&	X	\\
1975	&	12	07	41.02	&	$-$	29	49	18.0	&	5801	&	4	&	0	&	16	&	3.55	&	0	&	4.34	&	\nodata	\\
2026	&	12	17	41.99	&	+	28	02	53.1	&	7036	&	0	&	0	&	15	&	0	&	0	&	5.13	&	X	\\
2036	&	12	18	47.38	&	+	28	10	51.0	&	7036	&	3	&	5	&	16	&	3.13	&	2.69	&	4.72	&	X	\\
2282	&	12	55	01.06	&	+	11	29	33.8	&	4037	&	9	&	23	&	33	&	2.65	&	3.20	&	4.58	&	X	\\
2392	&	13	17	16.82	&	+	58	05	28.6	&	6879	&	0	&	0	&	19	&	0	&	0	&	4.51	&	\nodata	\\
2437	&	13	30	36.79	&	+	24	54	02.3	&	498	&	6	&	13	&	23	&	4.02	&	4.81	&	5.64	&	X	\\
2450	&	13	31	17.59	&	+	25	15	29.9	&	498	&	6	&	9	&	25	&	3.37	&	4.21	&	5.15	&	\nodata	\\
4740	&	13	31	28.42	&	+	25	00	59.4	&	498	&	7	&	9	&	9	&	3.88	&	4.14	&	3.77	&	\nodata	\\
2483	&	13	37	10.32	&	+	03	49	46.3	&	5547	&	2	&	5	&	12	&	3.04	&	3.14	&	4.19	&	\nodata	\\
4759	&	13	37	21.63	&	+	03	28	06.2	&	5547	&	3	&	6	&	14	&	3.04	&	3.39	&	4.88	&	\nodata	\\
2517	&	13	49	46.76	&	$-$	03	49	10.9	&	4261	&	9	&	12	&	24	&	4.33	&	4.52	&	4.74	&	\nodata	\\
2657	&	14	23	57.80	&	$-$	18	36	07.4	&	3454	&	0	&	8	&	18	&	0	&	2.53	&	4.29	&	\nodata	\\
2658	&	14	24	08.69	&	$-$	18	20	08.5	&	3454	&	7	&	14	&	23	&	3.37	&	2.91	&	4.22	&	\nodata	\\
4845	&	14	44	31.57	&	+	51	54	55.9	&	6317	&	3	&	6	&	18	&	2.98	&	2.97	&	4.83	&	X	\\
2748	&	14	48	26.28	&	$-$	16	26	27.2	&	3989	&	11	&	21	&	28	&	3.37	&	4.06	&	3.89	&	\nodata	\\
2799	&	15	04	52.79	&	$-$	33	08	13.1	&	6407	&	7	&	16	&	30	&	2.72	&	3.62	&	4.55	&	\nodata	\\
2858	&	15	17	41.03	&	+	22	44	24.8	&	8047	&	0	&	0	&	35	&	0	&	0	&	4.45	&	\nodata	\\
4917	&	15	33	40.36	&	+	31	27	40.5	&	7642	&	5	&	8	&	12	&	2.82	&	2.95	&	4.69	&	\nodata	\\
2923	&	15	33	56.47	&	+	31	17	33.2	&	7642	&	6	&	7	&	27	&	3.40	&	3.68	&	5.17	&	X	\\
2919	&	15	34	18.93	&	+	01	45	27.6	&	5708	&	4	&	0	&	29	&	2.60	&	0	&	4.53	&	\nodata	\\
2933	&	15	35	27.41	&	+	01	46	07.7	&	5708	&	9	&	14	&	18	&	3.80	&	4.78	&	4.91	&	\nodata	\\
2971	&	15	48	54.63	&	+	02	23	50.6	&	5397	&	5	&	10	&	24	&	3.09	&	2.91	&	4.42	&	\nodata	\\
2989	&	15	53	07.64	&	$-$	04	26	09.0	&	2911	&	5	&	7	&	15	&	3.10	&	2.87	&	4.11	&	\nodata	\\
5101	&	18	05	58.21	&	+	67	51	45.7	&	8780	&	0	&	0	&	17	&	0	&	0	&	4.61	&	X	\\
3439	&	18	53	01.06	&	+	59	20	28.2	&	4946	&	11	&	10	&	35	&	3.92	&	3.35	&	4.20	&	\nodata	\\
3564	&	21	36	17.10	&	+	00	59	14.3	&	7801	&	8	&	11	&	19	&	4.42	&	3.57	&	4.54	&	X	\\
5173	&	21	44	17.47	&	+	14	58	46.7	&	7605	&	0	&	4	&	12	&	0	&	2.78	&	4.52	&	\nodata	\\
5187	&	22	01	40.94	&	$-$	56	46	28.4	&	5652	&	5	&	7	&	22	&	3.15	&	2.53	&	4.28	&	X	\\
5188	&	22	02	06.59	&	$-$	56	37	04.1	&	5652	&	0	&	0	&	33	&	0	&	0	&	5.98	&	\nodata	\\
5192	&	22	04	18.62	&	$-$	56	40	10.4	&	5652	&	0	&	0	&	20	&	0	&	0	&	5.00	&	X	\\
5208	&	22	53	51.98	&	$-$	17	21	35.4	&	2074	&	12	&	19	&	33	&	3.78	&	4.23	&	3.60	&	X	\\
3739	&	23	14	31.50	&	$-$	42	32	24.9	&	5259	&	13	&	16	&	58	&	3.20	&	3.44	&	5.71	&	X	\\
3741	&	23	14	57.19	&	$-$	42	39	23.3	&	5259	&	0	&	0	&	45	&	0	&	0	&	4.82	&	\nodata	\\
5223	&	23	17	01.45	&	$-$	42	07	27.3	&	7582	&	17	&	33	&	49	&	4.00	&	4.36	&	5.07	&	X	\\
3752	&	23	19	30.71	&	$-$	36	04	13.2	&	7569	&	19	&	28	&	34	&	4.21	&	5.40	&	5.03	&	\nodata	\\
3765	&	23	28	07.20	&	$-$	29	59	04.5	&	4499	&	12	&	26	&	35	&	3.04	&	3.15	&	4.26	&	\nodata	\\
\enddata
\tablenotetext{a}{Count rates in the \eband~band, given in $10^{-3}$ sec$^{-1}$.}
\tablenotetext{b}{An entry of zero indicates a value less than the detection threshold limit of 2.5.}
\tablenotetext{C}{Identification of the source based on all the data in Tables
\ref{tab_nonxraydbrandom}, \ref{tab_xraydbrandom}, \& \ref{tab_2.1mrandom}:
C is a cluster of galaxies; X is not a cluster of galaxies; C: is a possible cluster of galaxies;  if there is no entry, no
definitive determination is possible at this time.}
\tablenotetext{d}{Source was removed from the Random Subsample, see \S \ref{subsec_databasesearch}, and
entry in Table
\ref{tab_nonxraydbrandom}.}
\end{\mydeluxetable}
\begin{deluxetable}{llcl}
\tablecaption{Identifications or Counterparts for Random Subsample Sources
\label{tab_nonxraydbrandom}}
\tablewidth{0pt}
\tablehead{
\colhead{Cat. \#} & \colhead{ID or Counterpart} & \colhead{Eval.\tablenotemark{a}} &
\colhead{References}
}
\startdata
64 & Radio: NVSS 002349.97 +171717; 17 mJy \\
97 & Radio: 87GB 00352.2+291502 &  & 1\\
   & Galaxy Group: NGC 0181/0183/0184 ($z = 0.018$) & C & 2\\
163 & QSO: 0050+0123 ($z = 1.439$) & A & 3\\
299 & Star: PPM 183386 (V = 10.0) & S &\\
    & Radio: PMN J0123--0348 &  & 4\\
1955 & AGN: EXO 1156.3+3239 ($z = 0.215$) & A & 5\\
2282 & Star: HD 112221 (GV, V = 9.10, B-V=0.45) & S & 6\\
4359\tablenotemark{b} & Galaxy Group related to A665 ($z=0.20$) &	C	& 7\\
5101 & Star: SAO 17771 (K2 V=9.3 B-V=1.4) & S & \\
5187 & IRAS: 21586--5652; probable AGN & A & 8\\
5208 & Star: EXO 2251.1--1737 (V=16) & S & 5\\
5223 & QSO: HB89 2314--423 ($z = 0.27$) & A & 9\\
5341 & Star: Gliese 1416 K2 double (V = 9.8, B-V = 1.2) & S & 10\\
\enddata
\tablenotetext{a}{Identification of the source based on the information available:
C is a cluster of galaxies; A is an AGN; S is a star.}
\tablenotetext{b}{Source was removed from the Random Subsample, due to being target related; see \S
\ref{subsec_databasesearch}.}
\tablerefs{\small{
(1) \citet{gre91}; (2) \citet{dre76}; (3) \citet{hew95};
(4) \citet{gri95}; (5) \citet{giom91}; (6) \citet{ols94}; (7) this paper; (8) \citet{mos90};
(9) \citet{hew87}; (10) \citet{cou94}
}
}
\end{deluxetable} 
\begin{deluxetable}{llrccclc}
\tablewidth{0pt}
\tabletypesize{\small}
\tablecaption{Other X-ray Detections of Random Subsample Sources\label{tab_xraydbrandom}}
\tablehead{
\colhead{Cat \#} & \colhead{Satellite/} & \colhead{\%\tablenotemark{b}} &
\colhead{$\Delta$\tablenotemark{c}} &
\colhead{3$\sigma$} & \colhead{Extent?} & \colhead{Comments} &
\colhead{Eval.\tablenotemark{e}} \\ 
& \colhead{Detector\tablenotemark{a}} & & [arcmin] & \colhead{Limit\tablenotemark{d}} & & &
}
\startdata
64 & R-H & $\geq$50\% & 0.7  &  & No & Point Source \\
& & & & & & Contributes & \\
69 & R-H & $\sim$100\% & 4.0 & 1.9 & No & & X \\ 
71 & R-H & 30-60\% & 0.9 & & No & Point Source  & \\
& & & & & & Contributes & \\
163 & R-H & $\geq$60\% & 0.7 & & No & QSO ID  & X \\
446 & R-P & 30-200\% & 0.9 & & No & Variable Source & X \\ 
657 & R-H, R-P & $\leq$50\% & 4.8 & 2.1 & No & & X \\
804 & R-H & $\leq$50\% & 1.7 & 1.0 & No & & \\
1415 & R-P & $\leq$30\% & 1.2 & & No & PSPC Source & \\
& & & & & & Contributes & \\
1424 & R-P & $\sim$100\% & 0.8 & & No & Other PSPC  & X \\
& & & & & & source nearby & \\
1477 & R-P & $\leq$50\% & 0.3 &  & ? & & \\
1757 & R-R & $\leq$30\% & 4.2 & & No & RASS Source  & \\
& & & & & & Contributes to & \\
& & & & & & 3rd aperture flux & \\
1918 & R-P & $\sim$50\% & 0.5 &  & ?  &  & \\
1955 & R-H & $\sim$100\% & 0.6 & & No & Two other HRI sources & X\\
 & & & & & & within 1' & \\
2036 & R-P & $\sim$100\% & 4.0 & 1.0 & ? & Contributes to 3rd & X \\
 & & & & & & aperture flux & \\
2437 & R-P & $\sim$100\% & 0.7 & & No & & X \\
2923 & R-P & $\sim$200\% & 0.9 & & No & Point Source & X \\
2933 & R-P & $\sim$100\% & 1.9 & 6.0 & ? & Short PSPC exposure & \\
2989 & R-P & $\geq$100\% & 0.4 & & ? & & \\
3564 & R-P & $\sim$100\% & 1.0 & & No & 2nd PSPC Source  & X \\
 & & & & & & 1\farcm3 away & \\
3739 & R-H, R-P & 80-150\% & 0.5 & & No & Variable Source & X \\
3741 & R-H, R-P & 10-30\% & 2.5 & 1.8 & No & Variable Source & P \\
3909 & R-H, R-P & $\sim$50\% & 1.2 & & No & Point Source & X \\
4057 & R-P & $\sim$50\% & 3.3 & 1.0 & No & Point Source & X \\
4374 & R-P & $\leq$50\% & 1.8 & 1.4 & ? & & \\
4508 & R-P & $\sim$50\% & 0.5 & & ?  & & \\
4845 & R-P & $\geq$50\% & 1.3 & 0.9 & ? & & \\
5187 & R-H, R-P & 200-400\% & 0.5 & & No & Variable Source:  & X \\
& R-R & & & & & ID=QSO & \\
5192 & R-P & $\sim$20\% & 4.1 & 0.7 & No & & X \\
5208 & R-P, E & $\sim$100\% & 0.6 & & No & ID=M8 Star & X \\
5223 & R-P & 80-100\% & 0.3 & & No & Variable Source:  & X \\
 & & & & & & ID=QSO & \\
\enddata
\tablenotetext{a}{R-H is {\it ROSAT} HRI, R-P is {\it ROSAT} PSPC, R-R is RASS, E is {\it
EXOSAT/CMA}.}
\tablenotetext{b}{Estimated percentage of the IPC flux
detected in the other dataset.}
\tablenotetext{c}{Angular offset between the positions of the source reported in this catalog
and the source in the other dataset in arcmin.}
\tablenotetext{d}{3 $\sigma$ flux limit within the 2\farcm5 diameter IPC detect aperture, in
units of
$10^{-13}$ \fx, where no source was detected in the first aperture.}
\tablenotetext{e}{Evaluation of the source based on the available X-ray data alone: X is not a
cluster of galaxies; C is a cluster of galaxies; P indicates that a point source contributes
significantly to the detected flux, but does not account for the entire source. No entry in
this column means that a definitive determination of source identification is not possible
based upon the X-ray data available.}
\end{deluxetable}
\begin{deluxetable}{llc}
\tablecaption{Optical Imaging of Sources from the Random Sample
\label{tab_2.1mrandom}}
\tablewidth{0pt}
\tablehead{
\colhead{Cat. \#} & 
\colhead{Comments} & \colhead{Eval.\tablenotemark{a}}
}
\startdata
64	&	Cirrus &	X	\\
69	&		&	X	\\
71	&		&	X	\\
97	&	Galaxy Group, NGC 181/183/184 ($z=0.018$)	&	C	\\
161	&	Possible cluster; B$_{gc} = 610-2230$~\bggmph; $z = 0.52-0.59$ & C: \\ 
& Second structure B$_{gc} = 370-1610$~\bggmph; $z=0.25-0.49$ \\ 
298	&		&	X	\\
302	&		&	X	\\
353	&		&	X	\\
446	&	Bright Star	&	X	\\
804	&		&	X	\\
1415	&		&	X	\\
1417	&	 &	X	\\
1477	&		&	X	\\
1568	&		&	X	\\
1584	&		&	X	\\
1587	&		&	X	\\
1757	&	Small group at $z\sim0.35$, but measured B$_{gc}<0$	&	X	\\
1955	&	{\it EXOSAT} QSO ($z=.215$)	&	X	\\
2026	&		&	X	\\
2036	&	Likely blend of 2 high-$z$ groups which have too & X \\
	& low a B$_{gc}$ value to account for the detection	&		\\
3886	&		&	X	\\
3915	&	PHL Object; likely AGN	&	X	\\
4040	&		&	X	\\
4057	&		&	X	\\
4374	&		&	X	\\
4375	&		&	X	\\
4845 &  & X \\
5341	&	Bright Star	&	X	\\
\enddata
\tablenotetext{a}{Evaluation of the source; letter designations are the same as in
Table
\ref{tab_randomsample}}
\end{deluxetable} 
\begin{deluxetable}{lrrccccrrr}
\tablecaption{X-ray data for Sources from the Ramp Subsample\label{tab_rampsample}}
\tablewidth{0pt}
\tablehead{
\colhead{Cat. \#} & \colhead{RA J2000} &
\colhead{Dec. J2000} & \colhead{Seq.} &
\colhead{$c.r._1$\tablenotemark{a}} &
\colhead{$c.r._2$\tablenotemark{a}} & \colhead{$c.r._3$\tablenotemark{a}} &
\colhead{$\sigma_1$} & \colhead{$\sigma_2$} &
\colhead{$\sigma_3$} 
}
\startdata
623	&	03	07	24.71	&	+	17	17	35.2	&	6830	&	4	&	7	&	12	&	4.36	&	4.38	&	5.56	\\
641	&	03	15	04.12	&	+	14	37	55.0	&	3954	&	5	&	5	&	9	&	5.22	&	5.15	&	3.47	\\
793	&	03	52	05.28	&	+	24	40	49.0	&	3175	&	24	&	33	&	45	&	4.23	&	4.41	&	5.16	\\
795	&	03	53	35.05	&	+	25	36	22.3	&	7408	&	10	&	16	&	28	&	4.19	&	4.18	&	4.60	\\
992	&	04	49	39.30	&	$-$	08	47	16.1	&	3748	&	6	&	12	&	15	&	3.42	&	4.14	&	4.87	\\
1033	&	05	04	30.95	&	$-$	11	51	52.1	&	10225	&	4	&	8	&	17	&	4.14	&	4.29	&	6.11	\\
1310	&	07	16	39.58	&	+	37	19	27.8	&	3554	&	7	&	9	&	25	&	2.87	&	2.79	&	4.01	\\
1328	&	07	38	31.78	&	+	65	24	08.3	&	589	&	4	&	9	&	18	&	2.66	&	3.26	&	4.91	\\
1342	&	07	47	12.51	&	+	38	48	11.8	&	3148	&	6	&	14	&	28	&	2.88	&	3.42	&	4.62	\\
1350	&	07	56	23.86	&	+	39	02	04.6	&	2622	&	11	&	12	&	37	&	3.46	&	2.95	&	4.18	\\
1492	&	08	51	40.06	&	+	33	31	23.3	&	3921	&	23	&	34	&	41	&	3.25	&	3.68	&	4.22	\\
1605	&	09	39	58.76	&	$-$	02	49	26.6	&	7427	&	7	&	15	&	13	&	3.64	&	4.69	&	4.22	\\
4045	&	02	35	04.04	&	$-$	03	40	44.5	&	7922	&	5	&	11	&	16	&	3.32	&	4.02	&	4.49	\\
\enddata
\tablenotetext{a}{Count rates are given in $10^{-3}$ sec$^{-1}$.}
\tablecomments{A few sources (e.g., \#641) only increase their S/N in the 4th aperture (data not shown), thereby
allowing them to be included in the ``Ramp'' subsample.}
\end{deluxetable}
\begin{deluxetable}{llc}
\tablecaption{Summary of Imaging and Database Search for Sources from the
Ramp Subsample
\label{tab_2.1mramp}}
\tablewidth{0pt}
\tablehead{
\colhead{Cat. \#} & 
\colhead{Comments} & \colhead{Eval.\tablenotemark{a}}
}
\startdata
623	&		& X	\\
641	&		2 bright stars $>4\arcmin$ offset & X	\\
793	&		Variable PSPC sources account for $100\%$ of flux & X \\
&  ID=K5 star & \\
795	&		V=6.3 mag star $4\farcm6$ N could contribute to 3rd aperture flux & X \\ 
992	&		Abell cluster 516 (z=0.141) & C \\
1033	&		 &	X \\
1310		&	B$_{gc} = 790-1410$~\bggmph; $z=0.34-0.40$	& C	\\
1328	&		{\it ROSAT} PSPC source contributes $20\%$ of flux & X	\\
& in 2nd aperture \\
1342	&		Bright star and FIRST radio source at $1\arcmin$ offset & X	\\
1350	&		several FIRST radio sources at $<1\arcmin$; & X \\
& no significant galaxy overdensity \\
1492	&		B$_{gc} = 1550-2920$~\bggmph; $z=0.42-0.50$	& C \\
& pair of BCGs and a blue arc \\
& RASS FSC detection 52\arcsec S of galaxy overdensity \\
& {\it ROSAT} source flux consistent with cluster ID \\
1605	&	B$_{gc} = 670-1050$~\bggmph; $z\approx 0.22-0.27$ & C \\
4045	&	Unresolved {\it ROSAT} X-ray source $0\farcm6$ offset & X	\\
& accounts for $100\%$ of flux \\
\\
\enddata
\tablenotetext{a}{Letter designations are the same as in Table \ref{tab_randomsample}.}
\end{deluxetable} 
\begin{deluxetable}{lrrccccrrr}
\tablecaption{X-ray Data for Sources Observed in the High Flux and S/N Subsample\label{tab_highsample}}
\tablewidth{0pt}
\tablehead{
\colhead{Cat. \#} & \colhead{RA J2000} &
\colhead{Dec. J2000} & \colhead{Seq.} &
\colhead{$c.r._1$\tablenotemark{a}} &
\colhead{$c.r._2$\tablenotemark{a}} & \colhead{$c.r._3$\tablenotemark{a}} &
\colhead{$\sigma_1$\tablenotemark{b}} & \colhead{$\sigma_2$\tablenotemark{b}} &
\colhead{$\sigma_3$\tablenotemark{b}} 
}
\startdata
1641	&	09	53	07.80	&	+	07	33	36.6	&	5934	&	7	&	19	&	21	&	4.13	&	6.08	&	6.14	\\		
1681	&	10	07	51.41	&	+	12	45	43.6	&	563	&	6	&	9	&	10	&	4.52	&	5.12	&	3.79	\\		
1767	&	10	48	28.84	&	+	06	43	26.8	&	9049	&	9	&	23	&	38	&	2.57	&	2.88	&	4.75	\\		
1772	&	10	49	48.98	&	+	06	44	53.6	&	9049	&	12	&	24	&	46	&	3.58	&	4.48	&	5.41	\\		
2128	&	12	31	56.36	&	+	62	35	32.8	&	6869	&	8	&	12	&	25	&	4.46	&	5.09	&	5.96	\\		
2407	&	13	19	36.59	&	+	55	06	35.6	&	4603	&	12	&	32	&	36	&	2.66	&	3.19	&	4.13	\\		
2436	&	13	30	11.38	&	+	30	43	51.3	&	491	&	9	&	14	&	21	&	4.52	&	5.66	&	6.91	\\		
2465	&	13	35	10.41	&	+	17	13	50.6	&	5376	&	15	&	33	&	39	&	3.97	&	5.44	&	5.87	\\		
2727	&	14	42	08.18	&	+	28	36	28.3	&	237	&	15	&	29	&	39	&	3.89	&	4.17	&	4.38	\\		
2844	&	15	12	27.90	&	+	72	00	47.6	&	6891	&	6	&	7	&	0	&	5.66	&	4.65	&	0	\\		
2906	&	15	31	55.38	&	+	24	20	42.9	&	3121	&	15	&	19	&	19	&	4.86	&	5.06	&	4.69	\\		
2937	&	15	36	02.20	&	+	23	18	34.3	&	10464	&	2	&	11	&	15	&	2.88	&	4.79	&	6.40	\\		
3065	&	16	08	53.06	&	+	28	53	31.3	&	5719	&	6	&	9	&	19	&	3.56	&	4.53	&	5.95	\\		
3068	&	16	10	07.13	&	+	19	05	43.3	&	3040	&	0	&	25	&	33	&	0	&	3.05	&	4.64	\\		
3244	&	17	08	12.15	&	+	54	43	48.1	&	7663	&	8	&	24	&	29	&	3.43	&	3.98	&	5.01	\\		
3299	&	17	29	35.78	&	+	52	30	46.1	&	3812	&	20	&	37	&	55	&	3.43	&	4.35	&	5.23	\\		
3353	&	17	55	51.73	&	+	67	53	44.9	&	8757	&	0	&	25	&	60	&	0	&	2.81	&	4.54	\\		
4725	&	13	19	16.19	&	$-$	12	23	16.6	&	10244	&	9	&	17	&	19	&	3.28	&	4.73	&	5.65	\\		
4746	&	13	34	03.27	&	$-$	08	26	14.4	&	917	&	6	&	10	&	16	&	3.77	&	4.84	&	5.44	\\		
5064	&	17	18	33.36	&	+	17	49	21.4	&	7481	&	5	&	10	&	15	&	3.31	&	5.11	&	3.36	\\		
\enddata
\tablenotetext{a}{Count rates are given in $10^{-3}$ sec$^{-1}$.}
\tablenotetext{b}{An entry of zero indicates a value less than 2.5.}
\end{deluxetable}
\begin{deluxetable}{lllc}
\tablecaption{Summary of Imaging and Database Search for Sources from the
High Flux and S/N Subsample
\label{tab_2.1mhigh}}
\tablewidth{0pt}
\tablehead{
\colhead{Cat. \#} 
& \colhead{Comments} & \colhead{ID\tablenotemark{a}}
}
\startdata
1641	&		&	X	\\
1681	&	Cluster $z	= 0.35-0.45$; B$_{gc}=180-1030$~\bggmph &	C	\\
1767	&	Cluster at $z=0.37-0.45$; B$_{gc}=880-1260$~\bggmph & X \\
& RASS FSC detection coincides with blue galaxy near IPC X-ray centroid \\
& {\it ROSAT} source flux matches 2nd IPC aperture flux \\
& Galaxy overdensity is in 3rd IPC aperture, likely blend with AGN\\
1772	&	Likely superposition of field galaxies; possible poor groups 	&	X	\\
2128	&	 	&	X	\\
2407	&	2 poor groups likely a blend with unidentified X-ray sources	&	X	\\
2436	&	Cluster $z=0.22-0.28$; B$_{gc}=1120-1680$~\bggmph & C \\
& possible 2nd cluster $5\arcmin$ NE, $z\sim0.30$; B$_{gc}=400-650$~\bggmph 	&	\\
2465 & & X \\
2727	&		&	X	\\
2844	&	Possible cluster $z=0.33-0.44$; B$_{gc}=640-1260$~\bggmph &	C:	\\
& RASS FSC detection at location of galaxy overdensity \\
& {\it ROSAT} source flux is consistent with cluster ID \\
2906	&	Possible cluster $z=0.42-0.53$; B$_{gc}=390-1410$~\bggmph	& C: \\
2937	&	Possible poor group at $z<0.1$; & X \\
& insufficient richness to account for the X-ray source	&	\\ 
3065	&	Possible blend of cluster $6\arcmin$ SW and unidentified sources	&	X	\\ 
3068	&		&	X	\\
3244	&		&	X	\\
3299	&		&	X	\\
3353	&	Definite $z<0.1$ group	&	C	\\
& RASS BSC detection just South of brightest E galaxy \\
& {\it ROSAT} source is extended, flux consistent with group ID \\
4725	&	 	&	X	\\
4746	&	Optical superposition of galaxies	&	X	\\
5064	&		&	X	\\
\enddata
\tablenotetext{a}{Evaluation of the source; letter designations are the same as
in Table
\ref{tab_randomsample}}\end{deluxetable} 
\begin{deluxetable}{lccccclc}
\tablecaption{New Clusters in the EMSS\label{tab_newclusters}}
\tablewidth{0pt}
\tablehead{
\colhead{Cat. \#}  & \colhead{$z$} & \colhead{B$_{gc}$\tablenotemark{a}} & \colhead{log L$_X$\tablenotemark{b}} &
\colhead{log L$_X$\tablenotemark{b}} &
\colhead{Sample} & \colhead{Notes} & \colhead{Ref.} \\ 
& & & from IPC & from
B$_{gc}$
\\ }
\startdata
97*	  & 0.018 & \nodata & $42.0\pm0.1$ & \nodata & Random & NGC 181/183  & (1) \\
& & & & & & /184 Group & (2) \\
161$^\dagger$  & $0.52-0.59$ & $610-2230$ & $44.7-45.0$ & $45.5$ & Random & Possible Cluster & (1) \\  
& & & & & & 2nd Cluster & (2) \\
& & & & & & in Field \\
324  & 0.085 & \nodata & $43.5\pm0.1$ & \nodata & HFS & Shakbazian 41;  & (3) \\
& & & & & & found in MSS1 \\
420  & 0.833 & \nodata & $45.0-45.6$ & 45.2\tablenotemark{c} & HFS & Found by  & (4) \\  
& & & & & & \rosat{} PSPC \\
962  & 0.152 & \nodata & $43.9\pm0.1$ & \nodata & HFS & Found in MSS1 & (3) \\
1310* & $0.34-0.40$ & $790-1410$ & $44.5-44.9$ & $45.3$ & Ramp & & (2) \\
1492* & $0.42-0.50$ & $1550-2910$ & $44.9-45.3$ & $46.2$\tablenotemark{d} & Ramp & & (2) \\
& & & & & HFS \\
1605 & $0.22-0.27$ & $670-1050$ & $43.9-44.2$ & $44.8$ & Ramp & & (2) \\
& & & & & HFS \\
1681 & $0.35-0.45$ & $180-1030$ & $44.1-44.6$ & $44.4$ & HFS & & (2) \\
2203 & 0.39 & \nodata & $44.8-45.0$ & $44.8$\tablenotemark{c} & HFS & & (6) \\
2436 & $0.22-0.28$ & $1120-1680$ & $44.1-44.4$ & $45.6$ & HFS & GHO Cluster & (2) \\ 
& & & & & & 2nd Cluster & (5) \\
& & & & & & in Field \\
2626 & 0.138 & \nodata & $44.1-44.3$ & $44.3$\tablenotemark{c} & HFS & & (6) \\ 
2844$^\dagger$ & $0.33-0.44$ & $640-1260$ & $44.3-44.8$ & $44.8$ & HFS & Possible Cluster & (2) \\
2906$^\dagger$ & $0.42-0.53$ & $390-1410$ & $44.6-45.0$ & $45.0$ & HFS & Possible Cluster & (2) \\
3353 & $<0.1$ & \nodata & $<44.0$ & \nodata & HFS & & (2) \\
3469 & 0.0152 & \nodata & $42.2\pm0.1$ & 41.9\tablenotemark{c} & \nodata & Abell S840 & (7) \\
4402 & 0.022  & \nodata & $41.9-42.0$ & \nodata & HFS & CfaN Group \#54; & (3) \\
& & & & & &  found in MSS1 & (8) \\
\enddata
\tablecomments{Clusters denoted with a * fell below the revised S/N limits described in \S \ref{subsec_moreacc}.
Clusters denoted with a $^\dagger$ are possible cluster identifications, pending further
investigation.}
\tablenotetext{a}{Galaxy over-density in units of \bggmph.}
\tablenotetext{b}{Log of the cluster X-ray luminosity in the \eband~band in \lxh. IPC luminosity is
calculated from the third aperture count rate, see
\S~\ref{subsec_sumnonrandom}; the range of values includes an estimate of the Poisson error on the detected
count rate. For luminosity estimates based on B$_{gc}$-values, see
\S~\ref{subsec_opt}.}
\tablenotetext{c}{Luminosity is taken from the Reference indicated in Column (8), converted to the
\eband~band, see entry in Appendix \ref{sec_appendix}.}
\tablenotetext{d}{This luminosity is likely too high to be physical; the B$_{gc}-L_X$ relationship we use here may be
suspect above B$_{gc}\sim 1500$~\bggmph.
}
\tablerefs{
(1) This Paper;
(2) \citet{P3};
(3) \citet{sto83};
(4) \citet{ebe00};
(5) \citet{gun86};
(6) \citet{per00}; \citet{rom00};
(7) \citet{vik98a};
(8) \citet{ram97}
}
\end{deluxetable}
\begin{deluxetable}{ccccccc}
\tablecaption{The Revised EMSS XLF Data\label{tab_emssbins}}
\tablewidth{0pt}
\tablehead{
\colhead{} & \multicolumn{6}{c}{Log L$_X$ ($0.3-3.5$~keV) [ergs s$^{-1}$] Luminosity Bins} \\
 & \colhead{$43.7-44.0$} & \colhead{$44.0-44.3$} &
\colhead{$44.3-44.6$} & \colhead{$44.6-44.9$} & \colhead{$44.9-45.2$} & \colhead{$45.2-45.5$} \\
\colhead{Redshift Shell} \\
\colhead{$z$} & \multicolumn{6}{c}{Number of Clusters in each Bin} \\
}
\startdata
     	      & 2         & 4           & 5         & 5           & \nodata    & \nodata  \\
$0.14-0.20$	& 3         & 4           & 5         & 5           & \nodata    & \nodata  \\
            & $3.8$     & 4           & 5         & 5           & \nodata    & \nodata  \\
\\
          	 & 2         & 8           & 6         & 4           & \nodata    & 1        \\
$0.20-0.30$ & 2.3       & 9.3         & 6.3       & 4           & \nodata    & 1        \\
            & $2.6-3.6$ & $10.3-14.2$ & $6.6-7.6$ & 4           & \nodata    & 1        \\
\\
            & \nodata   & 4           & 5         & 7           & 4          & \nodata  \\
$0.30-0.60$	& \nodata   & 4.4         & 5.9       & 8.9         & 4.3        & 1        \\
            & \nodata   & $4.7-5.9$   & $6.6-9.2$ & $10.3-12.0$ & $4.5-4.9$  & 1        \\
\\
            & \nodata   & \nodata     & \nodata   & 1           & 1          & \nodata  \\
$0.60-0.85$	& \nodata   & \nodata     & \nodata   & 1           & 2          & \nodata  \\
            & \nodata   & \nodata     & \nodata   & 1           & $2.8$      & \nodata  \\
\enddata
\tablecomments{The first entry in each bin is the original H92 value modified to
reflect the updates for newer work described in \S \ref{subsec_xlf}.
The second entry in each bin reflects the number of clusters  resulting from the addition of just
those 7.5 new clusters found thus far in our identification work from Table
\ref{tab_newclusters}. The third entry is the value resulting from the addition of
the estimated $13.1-24.9$ clusters projected by this work to be missing from the entire EMSS sky
area (including the extrapolation at the upper end of the range for clusters not yet found in the
HFS sample; see
\S~\ref{subsec_xlf}). Note that we have added a new redshift bin to accommodate the new high-$z$
clusters now observed.}
\end{deluxetable}
\end{document}